\documentclass[preprint,12pt]{elsarticle}

\usepackage{amsmath,amssymb}
\usepackage{booktabs}
\usepackage{graphicx}
\usepackage{subcaption}
\usepackage{xcolor}
\usepackage{microtype}
\usepackage[hidelinks]{hyperref}
\hypersetup{
    pdftitle={Physics-informed neural networks in the small-data regime: A controlled comparison with classical PDE surrogates},
    pdfauthor={Kirill Katsuba, Oleg Rudenko, Egor Bugaev, Alexander Tarakanov}
}
\usepackage{enumitem}
\usepackage{siunitx}
\usepackage{placeins}
\usepackage{float}
\usepackage{array}
\usepackage{longtable}
\usepackage[ruled,vlined,linesnumbered]{algorithm2e}

\graphicspath{{figures/}}
\journal{Journal of Computational Physics}

\newcommand{\Ld}{\mathcal{L}_{\mathrm{data}}}
\newcommand{\Libc}{\mathcal{L}_{\mathrm{ibc}}}
\newcommand{\Lpde}{\mathcal{L}_{\mathrm{pde}}}

\begin{document}
\begin{frontmatter}

\title{Physics-informed neural networks in the small-data regime: A controlled comparison with classical PDE surrogates}

\author[hse]{Kirill Katsuba\corref{cor1}}
\ead{kkatsuba@hse.ru}
\author[hse]{Oleg Rudenko}
\author[hse]{Egor Bugaev}
\author[hse,vk]{Alexander Tarakanov}
\cortext[cor1]{Corresponding author.}
\affiliation[hse]{
    organization={Faculty of Computer Science, HSE University},
    addressline={11 Pokrovsky Bulvar},
    city={Moscow},
    postcode={109028},
    country={Russia}
}
\affiliation[vk]{
    organization={VK},
    addressline={39 Leningradsky Prospekt, Building 79},
    city={Moscow},
    postcode={125167},
    country={Russia}
}

\begin{abstract}
\enlargethispage{\baselineskip}
Neural surrogates for engineering models are often trained on large collections of simulated solutions. We investigate whether physics-informed neural networks (PINNs) can achieve accurate predictions from scarce labelled observations by incorporating governing equations and prescribed initial and boundary conditions. We introduce hierarchical loss weighting (H-PINN) and its gradient-direction-aware extension (HO-PINN), which adjust the emphasis on physical constraints according to higher-priority losses. Hierarchical PINNs are compared with Vanilla PINN, tuned gradient boosting, support-vector regression, and Gaussian-process regression on a heat equation and three heterogeneous two-phase-flow benchmarks. Experiments use 20--1000 labelled states and ten paired seeds. Each label is one coordinate--state observation, not a complete simulation. At least one PINN variant achieves the highest mean primary-target \(R^2\) in each interpolation and equation-assisted forecasting comparison at 20 and 50 labels. The advantage for water saturation persists across all tested budgets. Hierarchical weighting reduces variability across runs in selected settings. These results demonstrate that known physics can support accurate neural surrogate predictions from very few solution observations, at substantially higher fitting cost than the tested classical regressors.
\end{abstract}


\end{frontmatter}

\section{Introduction}
\label{sec:introduction}

Many subsurface decisions require a model to be evaluated far more often than a single forward study. History matching tests many geological hypotheses against production data. Uncertainty quantification propagates uncertain rock and fluid properties through the model. Optimization repeatedly changes wells or controls. In each case, many parameter or control hypotheses lead to many high-fidelity flow solves \cite{Tang2020Surrogate,Tang2021Surrogate,Mo2019UQ}. Even when one simulation is tractable, the complete many-query workflow can be prohibitively expensive.

Porous-media flow is also a difficult surrogate problem. Permeability is heterogeneous and may vary over several orders of magnitude. Pressure is comparatively smooth but globally coupled through an elliptic equation. Saturation is transported through the heterogeneous velocity field and develops moving fronts whose location depends nonlinearly on mobility ratios, relative-permeability parameters, and the rock field \cite{Mo2019UQ,Wen2022UFNO}. This contrast motivates reporting pressure and saturation separately and using water saturation as the primary flow target.


Surrogates represent this response in different ways.
Projection-based reduced-order models approximate dynamics in
low-dimensional spaces \cite{Benner2015Reduction}, while learned
nonlinear representations encode solution fields on reduced
manifolds \cite{Fresca2021Reduction}.
Convolutional encoder--decoders predict spatial fields
\cite{Zhu2018Bayesian,Mo2019UQ}, and neural operators learn mappings
between input and output functions
\cite{Lu2021DeepONet,Li2021FNO,Wen2022UFNO}.
Here we study coordinate-to-state regression from scattered
observations. Gradient boosting, support-vector regression,
and Gaussian-process regression provide the non-neural baselines
\cite{chen2016xgboost,drucker1997support,rasmussen2006gaussian}.
This formulation compares methods at a fixed number of revealed
solution values rather than complete training fields.

Physics-informed learning offers a different trade-off. Instead of learning only from solution values, a PINN differentiates a neural representation and evaluates the governing equations at coordinates that require no reference solution \cite{Raissi2019,Karniadakis2021}. Initial and boundary conditions provide further constraints. The central question of this study is therefore not whether PINNs can solve the equations in principle. We ask at which label budgets and for which solution structures this additional physical information outweighs the optimization difficulty and fitting cost of PINNs.

Small data has a precise meaning in our protocol. A labelled state is one coordinate--state pair. For the flow problems it has the form
\begin{equation}
(t,x,y,\boldsymbol\mu)\longmapsto(p,S_w,S_o),
\end{equation}
where \(\boldsymbol\mu\) is absent in the fixed-parameter tests. Thus, \(N=500\) denotes 500 individual states, not 500 complete reservoir simulations, permeability realizations, or time-dependent fields. Every method receives the same \(N\) labels. The PINNs additionally receive the PDE, initial and boundary conditions, residual coordinates over the full time interval, and the permeability coefficient evaluated at residual points. Forecasting is consequently equation-assisted rather than purely data-driven.

The study makes four contributions. First, it compares six nested budgets from 20 to 1000 states using ten paired seeds and separate interpolation and forecasting regions. Second, it examines a smooth heat problem and heterogeneous two-phase-flow problems, including a seven-input parametric surrogate. Third, it compares a fixed PINN architecture with extensively tuned XGB, SVR, and GP baselines. Fourth, it introduces hierarchical exponential loss activation and its gradient-direction-aware extension, and assesses their accuracy and run-to-run reliability against unit-weight vanilla PINN. The experiments identify a PINN advantage at 20--50 labels and throughout the tested budgets for water saturation. They also identify settings where weighting reduces variability, settings where vanilla is more reliable, and the smooth, sufficiently sampled regime where GP is preferable.

\section{Related work}
\label{sec:related_work}

Data-driven porous-flow surrogates are commonly constructed from complete simulation realizations. Convolutional encoder--decoders have been used for porous-media
uncertainty quantification \cite{Zhu2018Bayesian} and dynamic
multiphase flow \cite{Mo2019UQ}. Recurrent and convolutional surrogates have also accelerated data assimilation and history matching in two- and three-dimensional reservoirs \cite{Tang2020Surrogate,Tang2021Surrogate}. These studies learn strong field-level representations but their training unit is typically a realization or a sequence of fields. Our training unit is instead one labelled state at one coordinate and time.

Operator-learning methods extend field prediction across coefficients and controls. DeepONet learns nonlinear operators from paired functions \cite{Lu2021DeepONet}. The Fourier neural operator performs global spectral mixing for parametric PDEs \cite{Li2021FNO}. U-FNO augments this construction with a U-Net path and has been applied to multiphase flow and geological carbon storage \cite{Wen2022UFNO}. Such models are designed for reusable maps between fields. The present study addresses the complementary setting in which only tens to hundreds of scattered state values are exposed to the learner while the differential operator is known.

PINNs incorporate this operator by automatic differentiation and penalize residuals at unlabelled coordinates \cite{Raissi2019,Karniadakis2021}. Sun et al. constructed fluid-flow surrogates without simulation labels by imposing conservation equations \cite{Sun2020}. He et al. combined multiphysics constraints with sparse observations for subsurface transport data assimilation \cite{He2020}. Chen et al. used physical constraints to identify governing equations from scarce measurements \cite{Chen2021}. For two-phase flow, residual adaptivity and finite-volume-informed formulations have been used to improve front resolution in heterogeneous media \cite{Hanna2022,Yan2024}. Coupled pressure and saturation studies nevertheless report persistent difficulty near sharp fronts \cite{Kenzhebek2025}.

Physics information does not remove numerical and optimization limitations. Standard PINNs can fail even on relatively simple PDEs \cite{Krishnapriyan2021}. Broad benchmarks show strong dependence on the equation, architecture, sampling, and optimizer \cite{Hao2024}. Finite elements can also be more accurate and efficient when the task is simply to solve one forward problem \cite{Grossmann2024}. We therefore keep the reference solver as ground truth, define the task as surrogate reconstruction after labels have been revealed, and report fitting cost rather than presenting PINNs as replacements for mature discretizations.

Loss balancing is a central part of reliable PINN training. Gradient-statistics annealing balances back-propagated gradients
across loss terms \cite{Wang2021Gradient}. Self-adaptive PINNs learn pointwise emphasis \cite{McClenny2023}. Inverse-Dirichlet weighting estimates imbalance from gradient statistics \cite{Maddu2022}, while relative-progress methods compare the rates at which objectives improve \cite{Bischof2025}. Our H-PINN instead activates IBC and PDE objectives as higher-priority losses approach prescribed scales. HO-PINN further reduces scalar weight for gradient directions already represented by earlier objectives. Unlike PCGrad \cite{Yu2020}, projection is used only to form weights and does not replace the gradient in the final backward pass.

The classical baselines provide complementary inductive biases.
Gradient boosting uses additive trees
\cite{friedman2001greedy,chen2016xgboost}.
Support-vector regression uses a regularized kernel representation
\cite{drucker1997support}. We use an RBF kernel in our experiments.
GP regression uses a covariance-based prior over functions
\cite{rasmussen2006gaussian}.
We examine when known physics improves prediction relative to
these tuned regressors at matched label budgets, and whether
hierarchical weighting improves run-to-run reliability.
Nested labels, paired seeds, and separate interpolation and
equation-assisted forecasting sets support this comparison.


\section{Small-data surrogate problem}
\label{sec:method}

\subsection{Learning task and architecture}

Let \(\boldsymbol z\) collect space, time, and any physical parameters, and let \(\boldsymbol y(\boldsymbol z)\) denote the solution. All models receive
\begin{equation}
\mathcal D_N=\{(\boldsymbol z_i,\boldsymbol y_i)\}_{i=1}^{N}.
\end{equation}
Here \(N\) counts labelled solution states rather than complete simulations or fields. In Test 1, \(\boldsymbol z=(t,x)\) and \(\boldsymbol y=T\). In the fixed-parameter flow tests, \(\boldsymbol z=(t,x,y)\), while Test 4 appends the four relative-permeability parameters. The flow target is \(\boldsymbol y=(p,S_w,S_o)\). PINNs additionally evaluate equation residuals at \(N_r\) unlabelled coordinates. These coordinates contain no reference-solver targets. Permeability is evaluated there as a known PDE coefficient and is not passed to the network as an input feature.

All tests use a gated residual multilayer perceptron with six hidden layers of width 128. Inputs are mapped to \([-1,1]\) using known domain bounds. Two learned tanh transforms are multiplied elementwise before the GELU hidden layers, with residual connections after every second layer and Xavier initialization. The heat model returns temperature. The flow model returns pressure, two saturations, and four auxiliary phase-velocity components. Sigmoid transformations and saturation normalization enforce
\begin{equation}
0\le\widehat S_w,\widehat S_o\le1,
\qquad \widehat S_w+\widehat S_o=1.
\end{equation}

\subsection{Loss components and hierarchical weighting}

The objective combines mean-square data, initial/boundary-condition (IBC), and PDE errors,
\begin{equation}
\mathcal L=w_d\Ld+w_b\Libc+w_r\Lpde,
\qquad w_d,w_b,w_r\ge0.
\label{eq:total_loss}
\end{equation}
Each epoch uses all \(N\) fixed labels and \(N_r=2N\) freshly sampled residual points. The data loss compares temperature in Test 1 and \((p,S_o,S_w)\) in Tests 2--4. Vanilla PINN uses constant weights \(w_d=w_b=w_r=1\), hence \(\mathcal L_{\rm vanilla}=\Ld+\Libc+\Lpde\). H-PINN and HO-PINN normalize their weights to sum to one. Architecture, prescribed physics, labels, residual-sampling stream, and training schedule are shared across the three variants. The objective-scale difference is part of this implementation and may interact with gradient clipping.

H-PINN activates the IBC and PDE terms as preceding errors approach prescribed scales. With \(\operatorname{sg}\) denoting stop-gradient, define
\begin{equation}
q_d=\frac{\operatorname{sg}(\Ld)}{L_d^0},
\qquad q_b=\frac{\operatorname{sg}(\Libc)}{L_b^0}.
\end{equation}
The exponential gates and weights are
\begin{equation}
a_d^{\mathrm H}=1,\quad
a_b^{\mathrm H}=e^{-q_d},\quad
a_r^{\mathrm H}=e^{-\max(q_d,q_b)},\quad
w_j^{\mathrm H}=\frac{a_j^{\mathrm H}}{\sum_k a_k^{\mathrm H}}.
\label{eq:h_weights}
\end{equation}
The target scales control the transition between objectives. We use \(L_d^0=0.03\) and \(L_b^0=0.003\).

HO-PINN also accounts for repeated gradient directions. Write \(\boldsymbol g_j=\nabla_\theta\mathcal L_j\) and
\begin{equation}
\mathcal Q_{\boldsymbol b}(\boldsymbol v)
=\boldsymbol v-\frac{\langle\boldsymbol v,\boldsymbol b\rangle}
{\|\boldsymbol b\|_2^2+\varepsilon}\boldsymbol b.
\end{equation}
The implementation removes projections sequentially and computes
\begin{equation}
\rho_b=\frac{\|\mathcal Q_{\boldsymbol g_d}(\boldsymbol g_b)\|_2}
{\|\boldsymbol g_b\|_2+\varepsilon},\qquad
\rho_r=\frac{\|\mathcal Q_{\boldsymbol g_b}
(\mathcal Q_{\boldsymbol g_d}(\boldsymbol g_r))\|_2}
{\|\boldsymbol g_r\|_2+\varepsilon}.
\label{eq:orthogonal_fractions}
\end{equation}
It then uses
\begin{equation}
\begin{aligned}
a_d^{\mathrm{HO}}&=1,&
a_b^{\mathrm{HO}}&=a_b^{\mathrm H}[c+(1-c)\rho_b],\\
a_r^{\mathrm{HO}}&=a_r^{\mathrm H}[c+(1-c)\rho_r],&
w_j^{\mathrm{HO}}&=\frac{a_j^{\mathrm{HO}}}{\sum_k a_k^{\mathrm{HO}}}.
\end{aligned}
\label{eq:ho_weights}
\end{equation}
We set \(c=0.25\). Repeated directions receive less weight, while the positive floor preserves every active objective. At \(c=1\), HO-PINN reduces to H-PINN. All gates and gradient-direction factors are detached before the backward pass. Algorithm~\ref{alg:hierarchical_training} summarizes
the training procedure for both variants. ~\ref{app:tuning} describes parameter selection.

\section{PDE benchmarks and reference solutions}
\label{sec:benchmarks}

\subsection{Periodic heat equation}

Test 1 uses
\begin{equation}
\partial_t T-\partial_{xx}T=0,\qquad
(t,x)\in[0,0.05]\times[0,1],
\label{eq:heat}
\end{equation}
with periodic boundaries and \(T(0,x)=2+\sin(2\pi x)+\cos(4\pi x)\). Its exact solution is
\begin{equation}
T(t,x)=2+e^{-4\pi^2t}\sin(2\pi x)+e^{-16\pi^2t}\cos(4\pi x).
\end{equation}
This benchmark evaluates smooth reconstruction without numerical reference error.

\subsection{Heterogeneous two-phase flow}

Tests 2--4 model incompressible, immiscible water--oil displacement on \(\Omega=[0,1]^2\), neglecting gravity and capillary pressure. Phase saturation \(S_\alpha\) is the fraction of pore volume occupied by phase \(\alpha\in\{w,o\}\). Conservation and closure give
\begin{equation}
\phi\,\partial_t S_\alpha+\nabla\cdot\boldsymbol u_\alpha=0,
\qquad S_w+S_o=1,
\label{eq:conservation}
\end{equation}
where \(\phi\) is porosity and \(\boldsymbol u_\alpha\) is Darcy flux. The phases share pressure \(p\) and obey
\begin{equation}
\boldsymbol u_\alpha=-K(\boldsymbol x)\lambda_\alpha\nabla p,
\qquad \lambda_\alpha=\frac{k_{r\alpha}}{\mu_\alpha},
\label{eq:darcy}
\end{equation}
with absolute permeability \(K\), relative permeability \(k_{r\alpha}\), and viscosity \(\mu_\alpha\). Corey-type laws are used \cite{corey1954interrelation},
\begin{equation}
k_{rw}=k_w^0 S_w^{p_w},\qquad k_{ro}=k_o^0 S_o^{p_o}.
\end{equation}
Adding the phase equations yields
\begin{equation}
\nabla\cdot[K(\lambda_w+\lambda_o)\nabla p]=0,\qquad
\boldsymbol u_t=-K(\lambda_w+\lambda_o)\nabla p.
\label{eq:pressure_equation}
\end{equation}
Equivalently, transport satisfies
\begin{equation}
\phi\,\partial_t S_\alpha+\nabla\cdot(f_\alpha\boldsymbol u_t)=0,
\qquad f_\alpha=\frac{\lambda_\alpha}{\lambda_w+\lambda_o}.
\label{eq:fractional_flow}
\end{equation}

We fix \(\phi=0.1\), \(\mu_w=1\), and \(\mu_o=3\). Initially \(S_w=0\) and \(S_o=1\). Pressure is imposed as \(p(t,0,y)=1\) and \(p(t,1,y)=0\). Both phases have zero normal flux on \(y=0,1\). At the inlet \(x=0\), zero oil flux specifies pure-water injection. Water inflow follows from the pressure difference. Outflow composition follows the interior upwind state.

The PINN predicts phase velocities directly. Its residual loss is the sample mean of the squared residual norm for
\begin{equation}
\boldsymbol r_\theta=
\begin{bmatrix}
\phi\,\partial_t\widehat S_w+\nabla\cdot\widehat{\boldsymbol u}_w\\
\phi\,\partial_t\widehat S_o+\nabla\cdot\widehat{\boldsymbol u}_o\\
\widehat{\boldsymbol u}_w+Kk_{rw}(\widehat S_w)\nabla\widehat p/\mu_w\\
\widehat{\boldsymbol u}_o+Kk_{ro}(\widehat S_o)\nabla\widehat p/\mu_o\\
\widehat S_w+\widehat S_o-1
\end{bmatrix}.
\label{eq:flow_pinn_residual}
\end{equation}
The IBC loss enforces the conditions above. The data loss averages the summed squared errors of \(p,S_o,S_w\).

Tests 2 and 3 use \((p_w,p_o,k_w^0,k_o^0)=(2,4,1,0.3)\). Their permeability contrasts are approximately 24 and \(9.0\times10^4\), respectively. Test 4 uses a third fixed heterogeneous field and 100 parameter combinations with
\begin{equation}
p_w\in[1.5,2.5],\quad p_o\in[3,5],\quad
k_w^0\in[0.9,1.1],\quad k_o^0\in[0.2,0.4].
\end{equation}
Its input is \((t,x,y,p_w,p_o,k_w^0,k_o^0)\). Permeability is supplied only as a PDE coefficient at residual points. Test 1 has two inputs and final time 0.05. Tests 2 and 3 have three inputs and final times 0.30 and 0.05. Test 4 has seven inputs and final time 0.20. The heat reference is analytical. Each flow reference uses a \(64\times64\) finite-volume grid. Tests 2 and 3 each contain one flow realization, while Test 4 contains 100 parameter combinations. Test 4 holds out coordinates, not entire parameter combinations, so it does not establish generalization to unseen parameter realizations. Temperature is the primary Test 1 target and water saturation is the primary flow target.

\subsection{Reference discretization}

The cell-centred finite-volume solver recomputes pressure at every adaptive transport substep. Harmonic face transmissibilities handle permeability jumps. Phase fluxes use upstream fractional flow, and explicit saturation updates are divided by cell pore volume \(\phi\Delta x\Delta y\), with a CFL limit of 0.2. Pressure and saturation are stored synchronously at 101 times including \(t=0\). Final times retain developed displacement fronts. The PINN residuals and reference solver enforce the same continuum equations through different discretizations.

\section{Experimental protocol}
\label{sec:protocol}

\subsection{Training and evaluation sets}

Budgets are \(N\in\{20,50,100,200,500,1000\}\), with seeds 101--110 controlling label selection and PINN initialization. Training sets are nested within each seed and shared by all methods. Labels occupy \(t\le0.7t_{\max}\) and are balanced over ten temporal bins.

Fixed, disjoint test sets measure interpolation at \(t\le0.7t_{\max}\) and forecasting at \(t>0.7t_{\max}\). Test 1 has 29,041 interpolation and 12,336 forecasting points. Each flow test uses 1,500 points per region. Test 4 draws these sets from a deterministic pool of 100,000 states covering all 100 parameter combinations.

PINN residuals cover the full time interval, excluding current labels and evaluation coordinates. They contain no reference values. Forecasting therefore uses knowledge of the future differential operator but no future solution labels. Initial and boundary conditions are prescribed physical information.

\subsection{Optimization and baseline tuning}

All PINNs use 50,000 epochs of Adam with initial learning rate \(3\times10^{-3}\), cosine decay to \(10^{-5}\), and gradient clipping at one. The checkpoint with the minimum balanced training loss is retained. A separate set of 512 unlabelled physics-validation points provides diagnostics, not checkpoint selection. The fixed architecture and weighting parameters were selected in staged development runs described in ~\ref{app:tuning}.

Classical models fit each output independently. XGB uses histogram-based squared-error boosting. SVR uses a radial-basis kernel with input and target scaling. GP uses standardized inputs and constant-scaled stationary kernels with white noise. At every budget and seed, 96 XGB configurations, 96 SVR configurations, and 20 GP kernel--regularization combinations are evaluated using three strictly forward-time folds. GP also optimizes kernel parameters by marginal likelihood with three restarts. Preprocessing is fitted inside each fold, and negative MSE determines the selected parameters. Neither evaluation set is used for tuning.

\subsection{Metrics}

For \(m\) evaluation points, we report
\begin{equation}
\mathrm{MSE}=\frac1m\sum_i(y_i-\widehat y_i)^2,\qquad
R^2=1-\frac{\sum_i(y_i-\widehat y_i)^2}
{\sum_i(y_i-\overline y)^2}.
\end{equation}
Metrics are computed per run, followed by the mean and sample standard deviation across ten seeds. Test 1 uses temperature. Flow tests report water saturation and pressure separately, with saturation defining the primary ranking. Error bars describe run-to-run variability, not confidence intervals. We assess predictive reliability within each fixed test--budget--region condition using the std ratio \(s_{\rm H}/s_{\rm vanilla}\) or \(s_{\rm HO}/s_{\rm vanilla}\). Smaller ratios indicate less sensitivity to the joint change in labels and initialization. We also inspect lower-tail scores and negative \(R^2\) evaluations. These diagnostics describe predictive variability rather than smoothness of the training loss. Conditions share seeds and nested labels, so counts across conditions are descriptive and not independent significance tests.

\section{Results}
\label{sec:results}

\begin{figure}[!htbp]
\centering
\includegraphics[width=0.96\linewidth]{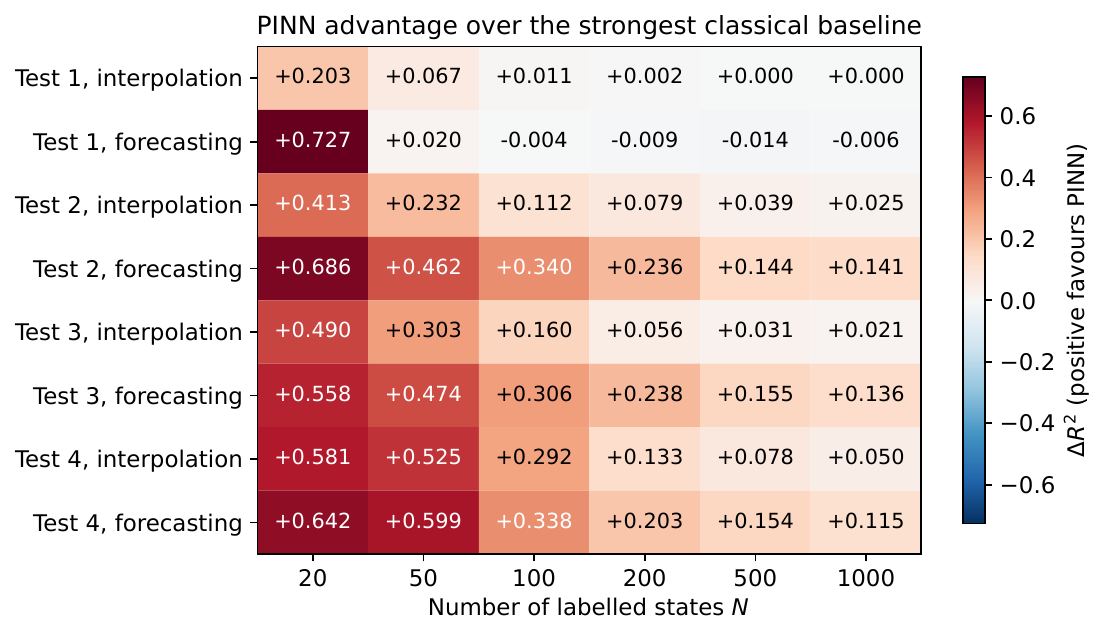}
\caption{Difference between the best PINN and best classical mean \(R^2\) for the primary target. Positive cells favour PINNs. Each value is computed from ten paired seeds. Forecasting uses future residual coordinates but no future solution labels.}
\label{fig:advantage_map}
\end{figure}

Table~\ref{tab:primary_r2} reports all six methods at the two smallest budgets. Figure~\ref{fig:advantage_map} extends the comparison to all six budgets by subtracting the strongest classical mean from the strongest of the three PINN means. The primary target is temperature in Test 1 and water saturation in Tests 2--4. Field figures show the fixed \(N=50\), seed-101 H-PINN and classical runs to illustrate spatial structure.
\begin{table}[!htbp]
\centering\fontsize{9}{11}\selectfont
\setlength{\tabcolsep}{0.5pt}
\caption{Primary $R^2\uparrow$ at 20 and 50 labels (mean $\pm$ std over ten seeds). Test 1 uses temperature and Tests 2--4 use water saturation. Vanilla denotes unit-weight PINN. Bold marks the highest mean.}
\label{tab:primary_r2}
\textbf{Interpolation}\par\smallskip
\begin{tabular}{@{}l@{\hspace{3pt}}r@{\hspace{5pt}}cccccc@{}}
\toprule
Test & $N$ & Vanilla & H-PINN & HO-PINN & XGB & SVR & GP \\
\midrule
T1 & 20 & $0.993\!\pm\!0.003$ & $1.000\!\pm\!0.000$ & {\bfseries\boldmath $1.000\!\pm\!0.000$} & $0.385\!\pm\!0.131$ & $0.461\!\pm\!0.303$ & $0.797\!\pm\!0.098$ \\
T1 & 50 & $0.997\!\pm\!0.001$ & {\bfseries\boldmath $1.000\!\pm\!0.000$} & $1.000\!\pm\!0.000$ & $0.639\!\pm\!0.117$ & $0.691\!\pm\!0.228$ & $0.932\!\pm\!0.055$ \\
\addlinespace[2pt]
T2 & 20 & {\bfseries\boldmath $0.954\!\pm\!0.007$} & $0.954\!\pm\!0.008$ & $0.937\!\pm\!0.052$ & $0.362\!\pm\!0.241$ & $0.396\!\pm\!0.276$ & $0.541\!\pm\!0.157$ \\
T2 & 50 & {\bfseries\boldmath $0.977\!\pm\!0.003$} & $0.975\!\pm\!0.004$ & $0.977\!\pm\!0.003$ & $0.632\!\pm\!0.046$ & $0.685\!\pm\!0.092$ & $0.745\!\pm\!0.089$ \\
\addlinespace[2pt]
T3 & 20 & {\bfseries\boldmath $0.755\!\pm\!0.018$} & $0.461\!\pm\!0.360$ & $0.567\!\pm\!0.312$ & $0.019\!\pm\!0.350$ & $0.084\!\pm\!0.482$ & $0.265\!\pm\!0.401$ \\
T3 & 50 & {\bfseries\boldmath $0.847\!\pm\!0.015$} & $0.841\!\pm\!0.015$ & $0.842\!\pm\!0.010$ & $0.344\!\pm\!0.302$ & $0.377\!\pm\!0.304$ & $0.544\!\pm\!0.247$ \\
\addlinespace[2pt]
T4 & 20 & $0.495\!\pm\!0.251$ & {\bfseries\boldmath $0.522\!\pm\!0.305$} & $0.426\!\pm\!0.358$ & $-0.058\!\pm\!0.268$ & $-0.187\!\pm\!0.294$ & $-0.077\!\pm\!0.339$ \\
T4 & 50 & $0.777\!\pm\!0.100$ & {\bfseries\boldmath $0.807\!\pm\!0.015$} & $0.779\!\pm\!0.109$ & $0.182\!\pm\!0.160$ & $0.083\!\pm\!0.109$ & $0.282\!\pm\!0.192$ \\
\bottomrule
\end{tabular}
\par\medskip
\textbf{Equation-assisted forecasting}\par\smallskip
\begin{tabular}{@{}l@{\hspace{3pt}}r@{\hspace{5pt}}cccccc@{}}
\toprule
Test & $N$ & Vanilla & H-PINN & HO-PINN & XGB & SVR & GP \\
\midrule
T1 & 20 & $0.946\!\pm\!0.069$ & $0.992\!\pm\!0.008$ & {\bfseries\boldmath $0.994\!\pm\!0.003$} & $-0.069\!\pm\!0.468$ & $-0.274\!\pm\!0.542$ & $0.267\!\pm\!0.817$ \\
T1 & 50 & $0.884\!\pm\!0.053$ & {\bfseries\boldmath $0.993\!\pm\!0.006$} & $0.993\!\pm\!0.005$ & $0.276\!\pm\!0.438$ & $0.479\!\pm\!0.524$ & $0.973\!\pm\!0.028$ \\
\addlinespace[2pt]
T2 & 20 & {\bfseries\boldmath $0.917\!\pm\!0.015$} & $0.917\!\pm\!0.016$ & $0.893\!\pm\!0.079$ & $0.230\!\pm\!0.287$ & $-0.864\!\pm\!1.240$ & $0.013\!\pm\!0.392$ \\
T2 & 50 & {\bfseries\boldmath $0.948\!\pm\!0.009$} & $0.947\!\pm\!0.010$ & $0.948\!\pm\!0.010$ & $0.397\!\pm\!0.136$ & $0.279\!\pm\!0.250$ & $0.486\!\pm\!0.197$ \\
\addlinespace[2pt]
T3 & 20 & {\bfseries\boldmath $0.642\!\pm\!0.042$} & $0.334\!\pm\!0.401$ & $0.445\!\pm\!0.365$ & $-0.089\!\pm\!0.269$ & $-2.148\!\pm\!4.703$ & $0.084\!\pm\!0.284$ \\
T3 & 50 & {\bfseries\boldmath $0.768\!\pm\!0.033$} & $0.762\!\pm\!0.026$ & $0.759\!\pm\!0.025$ & $0.172\!\pm\!0.394$ & $-1.011\!\pm\!2.289$ & $0.293\!\pm\!0.269$ \\
\addlinespace[2pt]
T4 & 20 & {\bfseries\boldmath $0.558\!\pm\!0.207$} & $0.550\!\pm\!0.360$ & $0.435\!\pm\!0.454$ & $-0.124\!\pm\!0.314$ & $-0.325\!\pm\!0.364$ & $-0.084\!\pm\!0.280$ \\
T4 & 50 & $0.794\!\pm\!0.059$ & {\bfseries\boldmath $0.815\!\pm\!0.011$} & $0.799\!\pm\!0.067$ & $0.198\!\pm\!0.269$ & $0.019\!\pm\!0.180$ & $0.216\!\pm\!0.214$ \\
\bottomrule
\end{tabular}
\par\medskip
\end{table}

Tables~\ref{tab:complete_test1}--\ref{tab:complete_test4} in ~\ref{app:complete_metrics} give primary-target \(R^2\) and MSE for
every budget and all six methods. Supplementary Tables S1--S3 provide the
corresponding pressure metrics for Tests 2--4.

\subsection{Test 1: periodic heat equation}

\begin{figure}[!htbp]
\centering
\includegraphics[width=\linewidth]{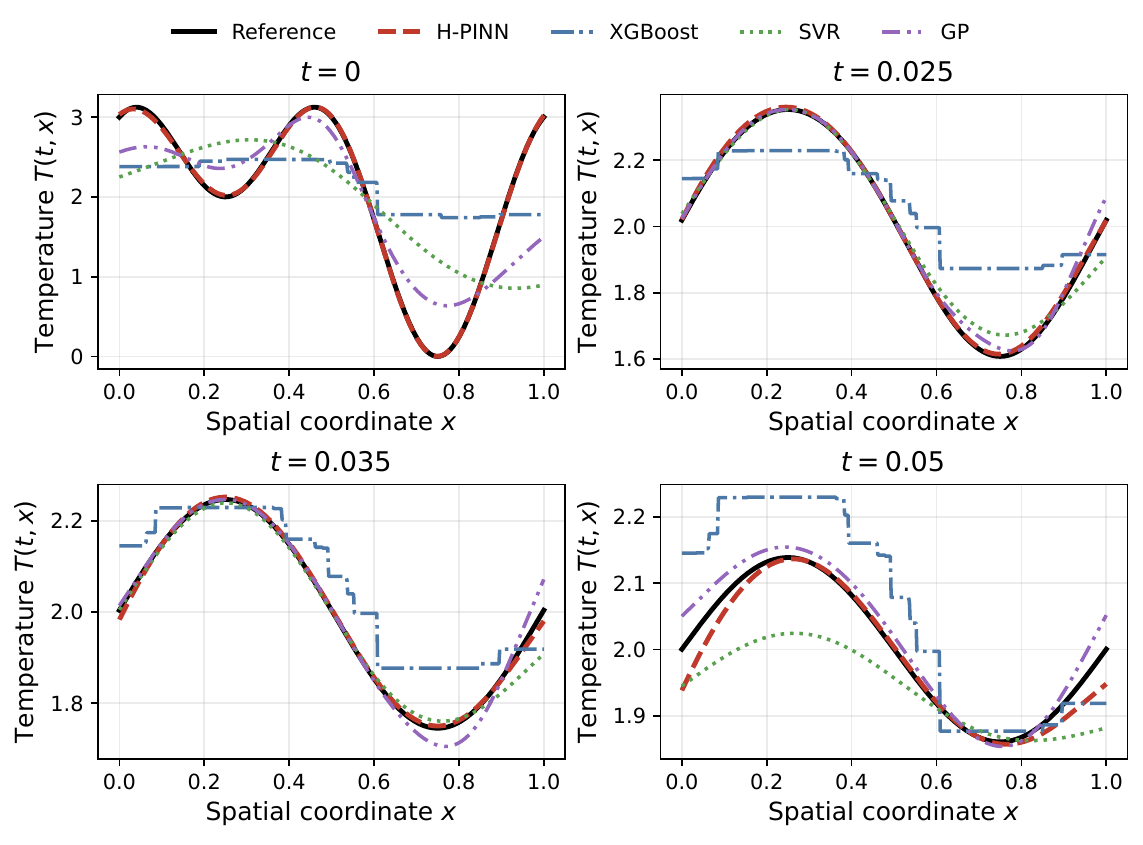}
\caption{Test 1 temperature for \(N=50\) and seed 101 at \(t=0,0.025,0.035,0.05\). Reference denotes the analytical solution.}
\label{fig:test1_fields}
\end{figure}

Figure~\ref{fig:test1_fields} compares temperature profiles at four times. H-PINN follows the analytical solution throughout the interval. Boosting predictions retain a piecewise-constant form, while SVR and GP smooth the sparsely sampled profile.

Both weighted PINNs are effectively exact in interpolation at \(N=20\) and 50. In forecasting, HO-PINN reaches \(0.994\pm0.003\) at \(N=20\), compared with \(0.946\pm0.069\) for vanilla and \(0.267\pm0.817\) for GP. GP leads forecasting from \(N=100\) and interpolation from \(N=500\), but its largest mean advantage over the best PINN is 0.014 in \(R^2\). This test identifies a regime in which enough labels make a classical smoothness prior sufficient.

\subsection{Test 2: moderate permeability contrast}

\begin{figure}[!htbp]
\centering
\includegraphics[width=\linewidth]{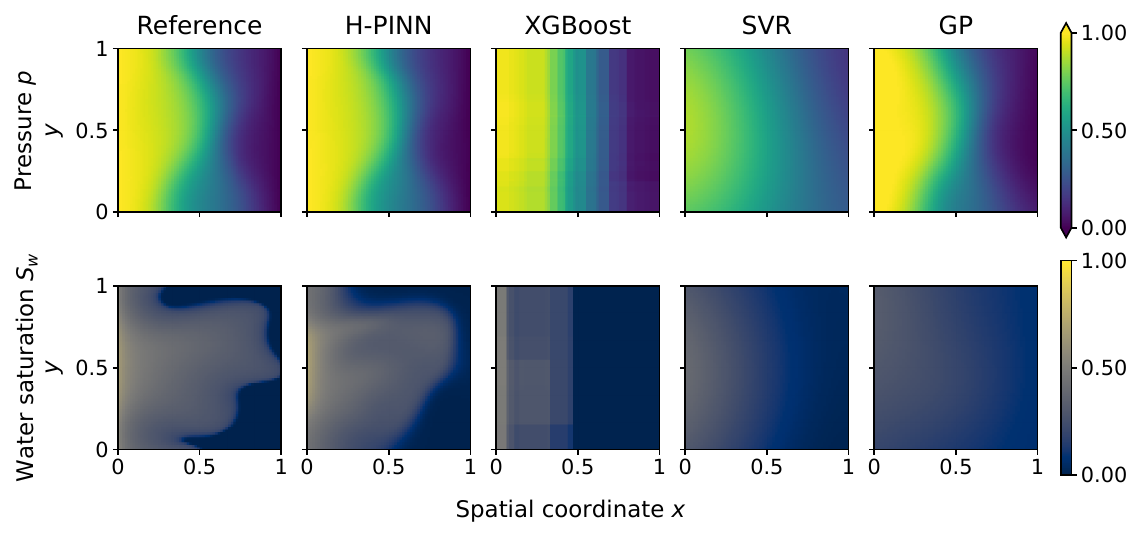}
\caption{Test 2 pressure and water saturation at \(t=0.30\), with \(N=50\) and seed 101. All predictions share the reference colour range within each row.}
\label{fig:test2_fields}
\end{figure}

The final-time fields in Fig.~\ref{fig:test2_fields} show the difference between the smooth pressure response and the transported saturation front. H-PINN preserves the front location and heterogeneous shape. The classical models recover the broad pressure trend but substantially smooth or discretize the saturation field.

At \(N=20\), H-PINN obtains saturation \(R^2=0.954\pm0.008\) in interpolation and \(0.917\pm0.016\) in forecasting. The strongest classical means are 0.541 and 0.230. At \(N=50\), the best PINN means increase to 0.977 and 0.948, with substantially smaller dispersion than the strongest classical entries. The advantage remains at \(N=1000\), where the best PINN forecasting mean is 0.993 and the best classical mean is 0.852.

\subsection{Test 3: high permeability contrast}

\begin{figure}[!htbp]
\centering
\includegraphics[width=\linewidth]{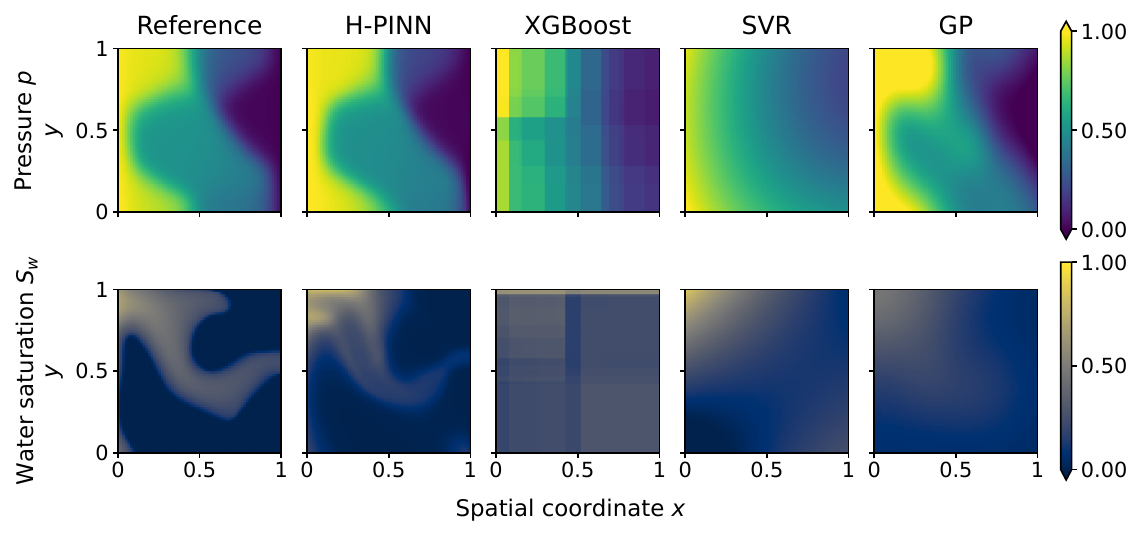}
\caption{Test 3 pressure and water saturation at \(t=0.05\), with \(N=50\) and seed 101.}
\label{fig:test3_fields}
\end{figure}

Test 3 increases the permeability contrast to approximately \(9.0\times10^4\). Figure~\ref{fig:test3_fields} shows that H-PINN reconstructs the main saturation pathway and pressure transition. XGB produces block artefacts, while SVR and GP miss much of the front geometry.

At \(N=20\), weighted PINNs have substantially greater seed variability than vanilla, although their means remain above the classical baselines. At \(N=50\), H-PINN reaches forecasting \(R^2=0.762\pm0.026\), compared with \(0.768\pm0.033\) for vanilla and \(0.293\pm0.269\) for GP. H-PINN and GP have mean MSE values of \(6.84\times10^{-3}\) and \(2.03\times10^{-2}\), respectively. The PINN advantage over classical models persists through \(N=1000\).

\subsection{Test 4: seven-input parametric surrogate}

\begin{figure}[!htbp]
\centering
\includegraphics[width=\linewidth]{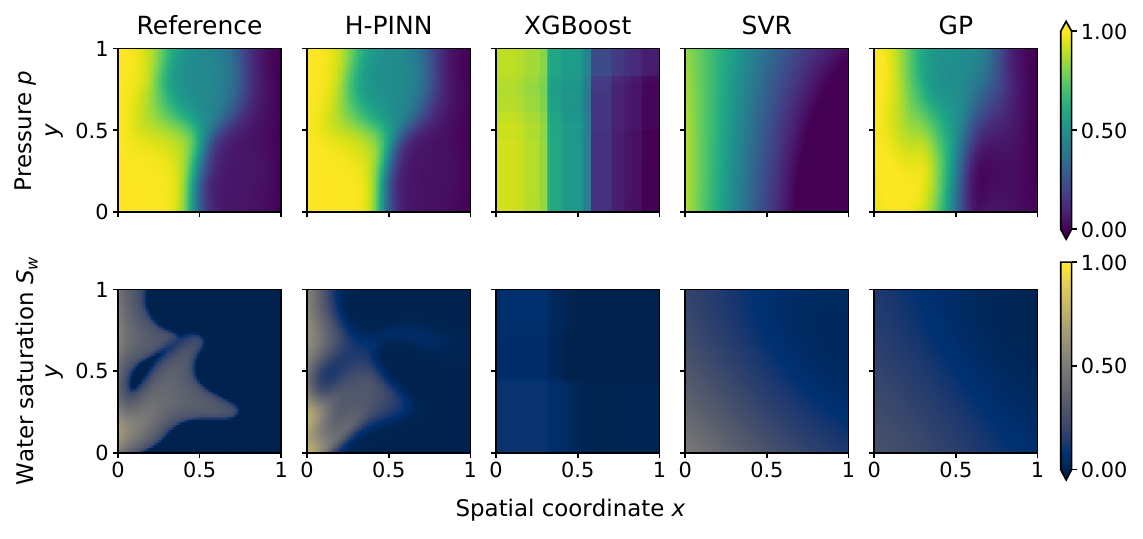}
\caption{Test 4 pressure and water saturation at \(t=0.20\) for parameter combination 7, with \(N=50\) and seed 101. The total label budget is shared across the pool of 100 combinations. No parameter combination is deliberately withheld.}
\label{fig:test4_fields}
\end{figure}

Test 4 trains one model across a pool of 100 relative-permeability parameter combinations. Figure~\ref{fig:test4_fields} displays one fixed combination at the final forecasting time. H-PINN follows the pressure interface and the principal saturation branches, while the three classical predictions lose much of the front structure.

The advantage persists in seven input dimensions. H-PINN forecasting saturation improves from \(0.550\pm0.360\) at \(N=20\) to \(0.815\pm0.011\) at \(N=50\). At the latter budget, vanilla gives \(0.794\pm0.059\), while the best classical mean is 0.216 with std 0.214. H-PINN therefore improves both the mean and dispersion in this condition. At \(N=20\), however, vanilla has smaller forecasting std than either weighted variant.

\subsection{Run-to-run reliability of loss weighting}

\begin{table}[!htb]
\centering\small
\caption{Run-to-run variability relative to vanilla PINN after 50,000-epoch training. Each test has 12 conditions (six budgets, two regions), each with ten paired seeds. Ratios below one favour weighting. Counts and medians are descriptive, not independent significance tests.}
\label{tab:stability}
\begin{tabular}{@{}lcccc@{}}
\toprule
& \multicolumn{2}{c}{Median std ratio $\downarrow$} & \multicolumn{2}{c}{Lower std, out of 12 $\uparrow$} \\
Test & H / Vanilla & HO / Vanilla & H-PINN & HO-PINN \\
\midrule
Test 1 & 0.19 & 0.16 & 11 & 12 \\
Test 2 & 1.11 & 1.14 & 4 & 3 \\
Test 3 & 0.98 & 0.95 & 7 & 6 \\
Test 4 & 1.03 & 1.01 & 5 & 6 \\
\bottomrule
\end{tabular}
\end{table}

Table~\ref{tab:stability} covers all budgets and both regions. H-PINN and HO-PINN each reduce std relative to vanilla in 27 of 48 conditions. The clearest benefit is Test 1, with median ratios 0.19 and 0.16. At \(N=50\), forecasting std decreases from 0.053 for vanilla to 0.006 for H and 0.005 for HO, while the mean also improves. In Test 4 at \(N=50\), H reduces forecasting std by a factor of 5.1 and raises the worst observed seed score from 0.632 to 0.795. HO does not reproduce this benefit in that condition.

The benefit is conditional. Vanilla has smaller std in most Test 2 conditions. In Test 3 at \(N=20\), its forecasting score is \(0.642\pm0.042\), whereas H and HO give \(0.334\pm0.401\) and \(0.445\pm0.365\). Across all 480 evaluations per PINN, H and HO have eight and seven negative primary \(R^2\) values, all in Tests 3--4 at \(N=20\), versus zero for vanilla. The two regions may flag the same trained model. Weighting can thus improve reliability markedly, but does not uniformly protect against low-budget failures. \ref{app:convergence} reports the separate fixed-epoch comparison.

\subsection{Cross-test result and computational cost}

At 20 and 50 labels, a PINN leads all 16 test--region--budget comparisons. Over \(N\le100\), a PINN leads in 23 of 24 comparisons. Across Tests 2--4, a PINN leads all 36 saturation comparisons, and across all tests it leads 42 of 48. The six classical wins are confined to the heat equation. Among all methods, vanilla leads 17 conditions, H leads 10 and HO leads 15.

This label efficiency has a substantial fitting cost. At \(N=50\), average GPU fitting times across the four tests range from 39--96 minutes for vanilla, 41--97 for H, and 64--158 for HO. Classical fitting including tuning takes approximately 0.8--66 seconds on CPU.

\section{Discussion}
\label{sec:discussion}

The ten-seed study supports a solution-label advantage under known governing equations and prescribed initial and boundary conditions. At 20--50 labels, a PINN leads every comparison with classical models. For water saturation, this advantage persists through 1000 labels. Pressure is smoother and generally easier to reconstruct, so averaging it with saturation would obscure front-prediction errors. Vanilla PINN demonstrates that much of the gain comes from physical supervision and the shared constrained architecture. Weighting adds a substantial reliability benefit for the heat problem and selected flow conditions, but can also suppress useful objectives for too long when labels are extremely sparse. 

The comparison holds solution labels fixed while allowing PINNs additional physical information. Residuals cover future times, which makes the forecasting task equation-assisted. It is not equivalent to unconstrained extrapolation from past labels. The results do not establish the same ranking when the PDE is unavailable, uncertain, or enforced only within the labelled interval.

Three limitations affect generalization. First, development inspected predictive metrics and reused seeds 101--103. Seeds 104--110 were added after configuration selection, but the protocol is not fully nested model selection. Second, the flow references use one spatial resolution and fixed permeability fields. Grid refinement and independent fields are needed to assess discretization sensitivity and transfer. Test 4 also does not hold out entire parameter combinations. Third, the comparison includes vanilla PINN and non-neural surrogates, but not specialized numerical solvers or other adaptive loss-weighting algorithms. The findings therefore concern these implementations rather than a ranking of weighting methods in general.

The heat-equation crossover shows that additional labels can make a conventional kernel surrogate preferable. PINN fitting is also substantially more expensive. The practical benefit depends on whether the reduction in required observations compensates for this cost. Our protocol establishes efficiency with respect to revealed state labels. It does not claim that \(N\) labels correspond to \(N\) full simulator runs. When sparse labels are extracted from already computed fields, reducing their number does not automatically reduce the offline simulation cost in the same proportion.

\section{Conclusions}

This study demonstrates that physics-informed neural networks can achieve accurate surrogate predictions from scarce labelled observations, including for strongly nonlinear, heterogeneous two-phase flow with moving saturation fronts. The benchmarks include permeability contrasts approaching five orders of magnitude and a seven-input parametric problem. Despite these challenges, a PINN achieves the highest mean primary-target accuracy among the tested methods in every interpolation and equation-assisted forecasting comparison at 20 and 50 labelled states. For water saturation, the advantage over tuned classical regression persists across all tested budgets up to 1000 states. These results show that useful neural surrogate predictions do not necessarily require large labelled datasets, even for challenging nonlinear flow problems, when reliable governing equations and prescribed initial and boundary conditions are available.

We introduce hierarchical loss weighting (H-PINN), which progressively activates physical constraints, and its gradient-direction-aware extension (HO-PINN). Both reduce variability across runs markedly for the heat equation and in selected flow settings. Their benefits remain problem dependent: neither consistently outperforms vanilla PINN, which is more reliable in some extremely sparse flow cases.

For engineering applications, the principal implication is that physical knowledge can supplement scarce solution observations. This label efficiency entails substantially higher fitting cost, and Gaussian-process regression becomes preferable for the sufficiently sampled smooth heat solution. Because each label represents one coordinate–state observation, the results establish efficiency with respect to revealed solution values rather than proportional savings in full simulations.

\FloatBarrier
\section*{Data and code availability}
The code and data are maintained in a single GitHub repository:
\begin{center}
\url{https://github.com/kirillkatsuba/pinn-small-data-jcp}
\end{center}

\section*{CRediT authorship contribution statement}

\textbf{Kirill Katsuba:}
Conceptualization, Methodology, Software, Investigation,
Formal analysis, Data curation, Validation, Visualization,
Writing -- original draft, Writing -- review \& editing.

\textbf{Oleg Rudenko:}
Conceptualization, Software, Validation,
Writing -- review \& editing.

\textbf{Egor Bugaev:}
Software, Validation, Writing -- review \& editing.

\textbf{Alexander Tarakanov:}
Conceptualization, Supervision,
Writing -- review \& editing.

\section*{Acknowledgments}
The supercomputer of HSE University was used to conduct the computational research~\cite{Kostenetskiy2026HPC}.

\section*{Declaration of competing interest}
The authors declare that they have no known competing financial interests or personal relationships that could have appeared to influence the work reported in this paper.

\section*{Funding}
This research did not receive any specific grant from funding agencies in the public, commercial, or not-for-profit sectors.

\section*{Declaration of generative AI and AI-assisted technologies in the writing process}
During preparation of this work, the authors used OpenAI Codex for code-assisted analysis, LaTeX drafting, and language editing. The authors reviewed and edited all generated material and take full responsibility for the content of the publication.

\bibliographystyle{elsarticle-num}
\renewcommand{\bibfont}{\footnotesize}
\setlength{\bibsep}{0pt}
\bibliography{references}

\FloatBarrier
\appendix
\counterwithin*{table}{section}
\renewcommand{\theHtable}{\Alph{section}.\arabic{table}}
\begingroup
\fontsize{9}{11}\selectfont
\section{Model selection and reproducibility}
\label{app:tuning}

\subsection{Staged PINN development}

Development used all four tests, \(N=50,200\), seeds 101--103, both regions
and 5000 epochs. Primary \(R^2\) was supplemented by pressure accuracy,
physics-validation losses, seed variability and fitting time. Selection was
sequential, first capacity, then activation, weighting scales, HO floor,
residual density and learning rate. Previously chosen settings were fixed.
Table~\ref{tab:pinn_screen} lists the candidates. Width 128 and depth six
provided the preferred accuracy--cost compromise. GELU had the highest
pooled development \(R^2\). This limited screen was not an exhaustive
architecture search. Both weighted variants were retained.

\begin{table}[!htb]
\centering\footnotesize
\caption{Sequential PINN development screen. Candidate sets on different
rows were not evaluated as one Cartesian product.}
\label{tab:pinn_screen}
\begin{tabular}{@{}p{0.27\linewidth}p{0.45\linewidth}p{0.18\linewidth}@{}}
\toprule
Component & Candidates & Selected\\
\midrule
Width, depth & \((64,4),(128,6),(128,8)\) & \((128,6)\)\\
Activation & tanh, SiLU, GELU & GELU\\
\(L_d^0\) & \(0.01,0.03,0.1,0.3\) & \(0.03\)\\
\(L_b^0\) & \(0.001,0.003,0.01,0.03\) & \(0.003\)\\
HO floor \(c\) & \(0,0.25,0.5,0.75,1\) & \(0.25\)\\
Residual ratio \(N_r/N\) & \(1,2,4\) & \(2\)\\
Initial learning rate & \(3\times10^{-4},3\times10^{-3},10^{-2}\) & \(3\times10^{-3}\)\\
\bottomrule
\end{tabular}
\end{table}

Selection used joint diagnostics without a preregistered scalar score.
The optimum in \(c\) was shallow around 0.25. Although \(L_b^0=0.001\)
slightly improved the initial pooled predictive score, \(0.003\) better
balanced IBC and PDE errors in later screens and was retained. Final
settings were shared across tests, budgets and seeds without retuning.

\subsection{Optimization and checkpointing}

Final runs use 50,000 Adam epochs, cosine decay from \(3\times10^{-3}\) to
\(10^{-5}\), and gradient-norm clipping at one. Adam was fixed after pilot
inspection. L-BFGS was excluded because fresh residuals change closure
objectives. Each epoch uses all \(N\) labels and \(2N\) fresh residual points
over the full horizon. Labels and evaluation coordinates are excluded from
the residual pool. The 512 fixed physics-validation points are diagnostic.
The checkpoint minimizes the balanced training objective, without predictive
or physics-validation epoch selection. Known domain bounds normalize inputs.
Initialization, labels and residual streams are paired across PINNs.
Of seeds 101--110, 101--103 were reused from development and 104--110 were
added after settings were frozen. Dispersion combines label and optimization
variability, with partial development-seed reuse acknowledged.

\begin{algorithm}[H]
\fontsize{9}{11}\selectfont
\SetAlgoNlRelativeSize{-1}
\SetNlSty{textnormal}{}{:}
\SetAlCapFnt{\bfseries\fontsize{9}{11}\selectfont}
\SetAlCapNameFnt{\normalfont\fontsize{9}{11}\selectfont}
\SetKwInput{KwData}{Data}
\SetKwInput{KwResult}{Result}
\SetKw{KwTo}{to}
\caption{Training with H-PINN and HO-PINN}
\label{alg:hierarchical_training}
\KwData{Labelled states $\mathcal D_N$; prescribed IBC and PDE; residual sampler;
variant $m\in\{\mathrm H,\mathrm{HO}\}$; epochs $E$; scales $L_d^0,L_b^0$; floor $c$.}
\KwResult{Selected network parameters $\theta_{\rm best}$.}
Initialize $\theta$, Adam, and the cosine learning-rate schedule; set $B\gets+\infty$\;
\For{$e\gets1$ \KwTo $E$}{
  Sample $2N$ residual coordinates over the full time horizon, excluding labels and evaluation states\;
  Evaluate $\Ld$, $\Libc$, and $\Lpde$ at the current $\theta$\;
  $q_d\gets\operatorname{clip}(\operatorname{sg}(\Ld)/(L_d^0+\varepsilon),0,M)$\;
  $q_b\gets\operatorname{clip}(\operatorname{sg}(\Libc)/(L_b^0+\varepsilon),0,M)$\;
  $a_d\gets1$, $a_b\gets\exp(-q_d)$, $a_r\gets\exp[-\max(q_d,q_b)]$\;
  \If{$m=\mathrm{HO}$}{
    $\boldsymbol g_j\gets\operatorname{sg}(\nabla_\theta\mathcal L_j)$ for $j\in\{d,b,r\}$\;
    $\boldsymbol v_b\gets\mathcal Q_{\boldsymbol g_d}(\boldsymbol g_b)$\;
    $\boldsymbol v_r\gets\mathcal Q_{\boldsymbol g_b}(\mathcal Q_{\boldsymbol g_d}(\boldsymbol g_r))$\;
    $\rho_b\gets\operatorname{clip}(\|\boldsymbol v_b\|_2/(\|\boldsymbol g_b\|_2+\varepsilon),0,1)$\;
    $\rho_r\gets\operatorname{clip}(\|\boldsymbol v_r\|_2/(\|\boldsymbol g_r\|_2+\varepsilon),0,1)$\;
    $a_b\gets a_b[c+(1-c)\rho_b]$, $a_r\gets a_r[c+(1-c)\rho_r]$\;
  }
  $w_j\gets\operatorname{sg}\bigl(a_j/(a_d+a_b+a_r+\varepsilon)\bigr)$ for $j\in\{d,b,r\}$\;
  $\mathcal L\gets w_d\Ld+w_b\Libc+w_r\Lpde$; $\ell\gets\operatorname{sg}(\mathcal L)$\;
  $\boldsymbol G\gets\nabla_\theta\mathcal L$ with $w_d,w_b,w_r$ held fixed\;
  Clip $\|\boldsymbol G\|_2$ to at most one and update $\theta$ with Adam\;
  Advance the cosine learning-rate schedule\;
  \If{$\ell<B$}{
    $B\gets\ell$; save the updated parameters as $\theta_{\rm best}$\;
  }
}
\Return{$\theta_{\rm best}$}\;
\end{algorithm}

\noindent
Here $\operatorname{sg}$ denotes stop-gradient, $\operatorname{clip}$ truncates
to the stated interval, and $\mathcal Q$ is the projection-removal operator in
Section~\ref{sec:method}. We use $\varepsilon=10^{-12}$ and $M=60$.
Projections determine scalar weights only. As in the training code, checkpoint
selection uses the pre-update objective and saves the corresponding
post-update parameters.

\subsection{Classical-model search}

Classical methods use the same labels, with one regressor per output.
Training times form four ordered blocks \(B_1,\ldots,B_4\), with folds

\begin{equation}
B_1\rightarrow B_2,\qquad
(B_1\cup B_2)\rightarrow B_3,\qquad
(B_1\cup B_2\cup B_3)\rightarrow B_4.
\end{equation}

Mean fold MSE selects the candidate, which is refitted on all \(N\) labels.
Scaling is fitted within each fold. Neither evaluation set enters tuning.

The procedure is not a greedy search. XGB and SVR use reproducible randomized
searches over fixed discrete spaces, while GP uses an exhaustive grid. XGB
evaluates 96 configurations per output, budget, and seed. Its space contains
\(n_{\rm trees}\in\{100,200,400,700,1000,1400,1800\}\), maximum depth in
\(\{2,3,4,5,6,8\}\), learning rate in
\(\{0.005,0.01,0.02,0.03,0.05,0.075,0.1\}\), child weight in
\(\{1,2,3,5,8,12\}\), row and feature fractions in
\(\{0.6,0.75,0.9,1\}\), split penalty in
\(\{0,10^{-3},10^{-2},0.1\}\), and \(L_1\) and \(L_2\) penalties in
\(\{0,10^{-3},10^{-2},0.1,1\}\) and
\(\{0.01,0.1,1,10,100\}\), respectively. Histogram-based squared-error
boosting is used.

RBF-SVR also evaluates 96 configurations. Inputs and targets are standardized
inside the cross-validation pipeline. The candidates are
\begin{align*}
C&\in\{0.03,0.1,0.3,1,3,10,30,100,300,1000,3000\},\\
\gamma&\in\{\text{scale},\text{auto},0.003,0.01,0.03,0.1,0.3,1,3,10,30\},\\
\epsilon&\in\{10^{-5},10^{-4},10^{-3},3\times10^{-3},10^{-2},0.03,0.05,0.1\}.
\end{align*}
The randomized candidate sequence is fixed by the experiment seed and output
index.

GP evaluates all 20 combinations of five covariance families and four nugget
values. The families are RBF, Mat\'ern with smoothness \(1/2\), \(3/2\), or
\(5/2\), and rational quadratic. Each is multiplied by a learned constant and
augmented by a learned white-noise term. The nugget is
\(\alpha\in\{10^{-10},10^{-8},10^{-6},10^{-4}\}\). For each candidate, kernel
amplitude, length scales, and noise level are optimized by marginal likelihood
with three restarts. Inputs are standardized and targets are normalized.
Neither the interpolation nor extrapolation evaluation set is accessed by any
of these searches.

\subsection{Reference-solution generation}

Test 1 uses the exact analytical solution at 161 times and 257 spatial
coordinates. Labels are sampled continuously across ten temporal bins.
Flow references use \(64\times64\) cell-centred grids and 101 synchronized
snapshots, including the initial state. Final times 0.30, 0.05 and 0.20
for Tests 2--4 accommodate different permeability scales and retain
developed fronts before outflow.

At every adaptive transport substep, pressure is recomputed from the current
total mobility. Face permeability is the harmonic mean of adjacent cells.
Phase transport uses first-order upstream fractional flow, and the conservative
update is divided by the cell pore volume. The step size obeys CFL \(=0.2\).
Generation aborts if saturation leaves \([0,1]\) beyond numerical tolerance or
if the pressure solve exceeds its residual tolerance. The stored runs have
pressure-balance errors below \(5\times10^{-14}\) and phase mass-balance errors
below \(2\times10^{-16}\).

Fixed permeability fields have contrasts of about 24 and \(9.0\times10^4\)
in Tests 2--3. Test 4 uses a third field and 100 fixed parameter tuples.
All fields and tuples are shared across methods. Training candidates cover
the first 70\% of time, with 100,000 pooled states for Test 4. Temporal
interleaving yields nested budgets. Evaluation states are removed before
label or residual sampling. Flow PINNs satisfy saturation bounds and have
closure errors near \(10^{-8}\), which do not determine predictive rankings.

\subsection{Fixed-epoch convergence comparison}
\label{app:convergence}

An additional 720 PINN fits cover all four tests, six budgets, ten seeds and
three variants. They use 10,000 epochs and a cosine horizon of 10,000 epochs,
with checkpoints at 1000, 2500, 5000, 7500 and 10,000. These are independent
short-budget runs, not prefixes of the main 50,000-epoch schedule. At epoch
1000, H and HO exceed vanilla in mean primary \(R^2\) in 24 and 23 of the
48 conditions. At epoch 10,000 these counts are 26 and 24, while smaller
standard deviations occur in 25 and 23 conditions. Evaluation uses exact
epochs, not selected checkpoints. These mixed rankings do not establish
universally faster convergence or a wall-clock advantage, particularly
because HO requires additional gradient evaluations.

\subsection{Exploratory architecture and hierarchy ablations}
\label{app:ablations}

CPU pilots at \(N=20\), seed 101 and 10,000 epochs remove all output
transformations while retaining the backbone, optimizer and losses.
Against exact final checkpoints of the constrained GPU controls, forecasting
saturation \(R^2\) decreases for every variant in Tests 2--3. For H-PINN it
falls from 0.862 to 0.845 and from 0.588 to 0.470, respectively. All direct
Test 4 runs have nonfinite weights by epoch 1000. Negative saturation raised
to a noninteger Corey exponent is undefined in real arithmetic, making
unrestricted outputs incompatible with the unmodified residual outside
the physical domain. Single-seed, cross-device pilots cannot establish
multi-seed robustness.

A second H-PINN pilot prioritizes IBC, PDE, then data. With
\(q_b=\operatorname{sg}(\Libc)/0.003\) and
\(q_r=\operatorname{sg}(\Lpde)/0.003\), its normalized weights are proportional
to \((a_b,a_r,a_d)=(1,e^{-q_b},e^{-\max(q_b,q_r)})\).
It gives no consistent improvement across Tests 1--4 and a final data weight
of only \(9\times10^{-4}\) in Test 3. We retain data--IBC--PDE priority.

\section{Complete primary-target metrics across label budgets}
\label{app:complete_metrics}

Tables~\ref{tab:complete_test1}--\ref{tab:complete_test4} report all six methods
at \(N=20,50,100,200,500,1000\). Interpolation and equation-assisted forecasting
are reported separately. Test 1 uses temperature and Tests 2--4 use water
saturation. Supplementary Tables S1--S3 give pressure metrics for the same
budgets and methods. Every cell uses the same ten paired seeds as
the main comparison. Bold identifies the best unrounded mean within each row,
so equally rounded entries need not share highlighting. All std values remain
visible without separate highlighting. MSE mantissas use one decimal place
and \(R^2\) uses three.

\begingroup
\fontsize{8}{10}\selectfont
\setlength{\tabcolsep}{0.5pt}
\renewcommand{\arraystretch}{0.98}
\setlength{\LTpre}{5pt}
\setlength{\LTpost}{5pt}
\begin{longtable}{@{}r@{\hspace{5pt}}cccccc@{}}
\caption{Test 1 temperature, all budgets (mean $\pm$ sample std over ten seeds). Bold marks the best unrounded mean. MSE entries use $(a\pm b)10^e$.}
\label{tab:complete_test1} \\
\toprule
$N$ & Vanilla & H-PINN & HO-PINN & XGB & SVR & GP \\
\midrule
\endfirsthead
\multicolumn{7}{c}{Table \thetable\ continued} \\
\toprule
$N$ & Vanilla & H-PINN & HO-PINN & XGB & SVR & GP \\
\midrule
\endhead
\midrule
\multicolumn{7}{r}{Continued on next page} \\
\endfoot
\bottomrule
\endlastfoot
\multicolumn{7}{@{}l}{\textit{Temperature, $R^2\uparrow$ --- Interpolation}} \\*
20 & $0.993\!\pm\!0.003$ & $1.000\!\pm\!0.000$ & {\bfseries\boldmath $1.000\!\pm\!0.000$} & $0.385\!\pm\!0.131$ & $0.461\!\pm\!0.303$ & $0.797\!\pm\!0.098$ \\
50 & $0.997\!\pm\!0.001$ & {\bfseries\boldmath $1.000\!\pm\!0.000$} & $1.000\!\pm\!0.000$ & $0.639\!\pm\!0.117$ & $0.691\!\pm\!0.228$ & $0.932\!\pm\!0.055$ \\
100 & $0.999\!\pm\!0.001$ & $1.000\!\pm\!0.000$ & {\bfseries\boldmath $1.000\!\pm\!0.000$} & $0.684\!\pm\!0.152$ & $0.777\!\pm\!0.222$ & $0.989\!\pm\!0.009$ \\
200 & $0.999\!\pm\!0.000$ & $1.000\!\pm\!0.000$ & {\bfseries\boldmath $1.000\!\pm\!0.000$} & $0.766\!\pm\!0.123$ & $0.971\!\pm\!0.071$ & $0.998\!\pm\!0.001$ \\
500 & $1.000\!\pm\!0.000$ & $1.000\!\pm\!0.000$ & $1.000\!\pm\!0.000$ & $0.862\!\pm\!0.124$ & $0.908\!\pm\!0.146$ & {\bfseries\boldmath $1.000\!\pm\!0.000$} \\
1000 & $1.000\!\pm\!0.000$ & $1.000\!\pm\!0.000$ & $1.000\!\pm\!0.000$ & $0.927\!\pm\!0.063$ & $0.891\!\pm\!0.130$ & {\bfseries\boldmath $1.000\!\pm\!0.000$} \\
\addlinespace[2pt]
\multicolumn{7}{@{}l}{\textit{Temperature, $R^2\uparrow$ --- Equation-assisted forecasting}} \\*
20 & $0.946\!\pm\!0.069$ & $0.992\!\pm\!0.008$ & {\bfseries\boldmath $0.994\!\pm\!0.003$} & $-0.069\!\pm\!0.468$ & $-0.274\!\pm\!0.542$ & $0.267\!\pm\!0.817$ \\
50 & $0.884\!\pm\!0.053$ & {\bfseries\boldmath $0.993\!\pm\!0.006$} & $0.993\!\pm\!0.005$ & $0.276\!\pm\!0.438$ & $0.479\!\pm\!0.524$ & $0.973\!\pm\!0.028$ \\
100 & $0.937\!\pm\!0.051$ & $0.993\!\pm\!0.008$ & $0.994\!\pm\!0.005$ & $0.409\!\pm\!0.404$ & $0.527\!\pm\!0.430$ & {\bfseries\boldmath $0.998\!\pm\!0.002$} \\
200 & $0.947\!\pm\!0.033$ & $0.990\!\pm\!0.007$ & $0.987\!\pm\!0.011$ & $0.707\!\pm\!0.241$ & $0.779\!\pm\!0.175$ & {\bfseries\boldmath $0.999\!\pm\!0.001$} \\
500 & $0.965\!\pm\!0.024$ & $0.986\!\pm\!0.023$ & $0.984\!\pm\!0.020$ & $0.844\!\pm\!0.146$ & $0.848\!\pm\!0.120$ & {\bfseries\boldmath $1.000\!\pm\!0.000$} \\
1000 & $0.982\!\pm\!0.019$ & $0.990\!\pm\!0.008$ & $0.994\!\pm\!0.004$ & $0.881\!\pm\!0.035$ & $0.738\!\pm\!0.246$ & {\bfseries\boldmath $1.000\!\pm\!0.000$} \\
\addlinespace[2pt]
\multicolumn{7}{@{}l}{\textit{Temperature, $\mathrm{MSE}\downarrow$ --- Interpolation}} \\*
20 & $(1.5\!\pm\!0.6)10^{-3}$ & $(1.0\!\pm\!0.8)10^{-4}$ & {\bfseries\boldmath $(9.9\!\pm\!6.7)10^{-5}$} & $(1.3\!\pm\!0.3)10^{-1}$ & $(1.2\!\pm\!0.7)10^{-1}$ & $(4.4\!\pm\!2.1)10^{-2}$ \\
50 & $(5.8\!\pm\!3.2)10^{-4}$ & {\bfseries\boldmath $(6.5\!\pm\!4.2)10^{-5}$} & $(6.8\!\pm\!4.1)10^{-5}$ & $(7.8\!\pm\!2.5)10^{-2}$ & $(6.7\!\pm\!4.9)10^{-2}$ & $(1.5\!\pm\!1.2)10^{-2}$ \\
100 & $(3.1\!\pm\!2.4)10^{-4}$ & $(4.8\!\pm\!3.1)10^{-5}$ & {\bfseries\boldmath $(4.4\!\pm\!2.4)10^{-5}$} & $(6.9\!\pm\!3.3)10^{-2}$ & $(4.8\!\pm\!4.8)10^{-2}$ & $(2.4\!\pm\!2.0)10^{-3}$ \\
200 & $(1.7\!\pm\!0.4)10^{-4}$ & $(3.8\!\pm\!1.3)10^{-5}$ & {\bfseries\boldmath $(3.6\!\pm\!1.6)10^{-5}$} & $(5.1\!\pm\!2.7)10^{-2}$ & $(6.4\!\pm\!15.5)10^{-3}$ & $(3.8\!\pm\!2.8)10^{-4}$ \\
500 & $(1.1\!\pm\!0.3)10^{-4}$ & $(3.6\!\pm\!3.3)10^{-5}$ & $(3.0\!\pm\!2.7)10^{-5}$ & $(3.0\!\pm\!2.7)10^{-2}$ & $(2.0\!\pm\!3.2)10^{-2}$ & {\bfseries\boldmath $(1.6\!\pm\!2.2)10^{-5}$} \\
1000 & $(6.1\!\pm\!2.1)10^{-5}$ & $(2.6\!\pm\!1.3)10^{-5}$ & $(2.0\!\pm\!0.6)10^{-5}$ & $(1.6\!\pm\!1.4)10^{-2}$ & $(2.4\!\pm\!2.8)10^{-2}$ & {\bfseries\boldmath $(1.6\!\pm\!1.7)10^{-6}$} \\
\addlinespace[2pt]
\multicolumn{7}{@{}l}{\textit{Temperature, $\mathrm{MSE}\downarrow$ --- Equation-assisted forecasting}} \\*
20 & $(9.8\!\pm\!12.6)10^{-4}$ & $(1.5\!\pm\!1.4)10^{-4}$ & {\bfseries\boldmath $(1.1\!\pm\!0.6)10^{-4}$} & $(1.9\!\pm\!0.9)10^{-2}$ & $(2.3\!\pm\!1.0)10^{-2}$ & $(1.3\!\pm\!1.5)10^{-2}$ \\
50 & $(2.1\!\pm\!1.0)10^{-3}$ & {\bfseries\boldmath $(1.3\!\pm\!1.1)10^{-4}$} & $(1.3\!\pm\!0.9)10^{-4}$ & $(1.3\!\pm\!0.8)10^{-2}$ & $(9.5\!\pm\!9.5)10^{-3}$ & $(4.9\!\pm\!5.1)10^{-4}$ \\
100 & $(1.1\!\pm\!0.9)10^{-3}$ & $(1.2\!\pm\!1.4)10^{-4}$ & $(1.1\!\pm\!0.9)10^{-4}$ & $(1.1\!\pm\!0.7)10^{-2}$ & $(8.6\!\pm\!7.8)10^{-3}$ & {\bfseries\boldmath $(3.3\!\pm\!3.5)10^{-5}$} \\
200 & $(9.6\!\pm\!5.9)10^{-4}$ & $(1.8\!\pm\!1.3)10^{-4}$ & $(2.4\!\pm\!2.0)10^{-4}$ & $(5.3\!\pm\!4.4)10^{-3}$ & $(4.0\!\pm\!3.2)10^{-3}$ & {\bfseries\boldmath $(1.5\!\pm\!1.1)10^{-5}$} \\
500 & $(6.4\!\pm\!4.4)10^{-4}$ & $(2.5\!\pm\!4.1)10^{-4}$ & $(2.9\!\pm\!3.7)10^{-4}$ & $(2.8\!\pm\!2.7)10^{-3}$ & $(2.8\!\pm\!2.2)10^{-3}$ & {\bfseries\boldmath $(2.3\!\pm\!1.4)10^{-6}$} \\
1000 & $(3.2\!\pm\!3.4)10^{-4}$ & $(1.8\!\pm\!1.5)10^{-4}$ & $(1.1\!\pm\!0.7)10^{-4}$ & $(2.2\!\pm\!0.6)10^{-3}$ & $(4.8\!\pm\!4.5)10^{-3}$ & {\bfseries\boldmath $(1.8\!\pm\!1.1)10^{-6}$} \\
\end{longtable}
\endgroup

\begingroup
\fontsize{8}{10}\selectfont
\setlength{\tabcolsep}{0.5pt}
\renewcommand{\arraystretch}{0.98}
\setlength{\LTpre}{5pt}
\setlength{\LTpost}{5pt}
\begin{longtable}{@{}r@{\hspace{5pt}}cccccc@{}}
\caption{Test 2 water saturation, all budgets (mean $\pm$ sample std over ten seeds). Bold marks the best unrounded mean. MSE entries use $(a\pm b)10^e$.}
\label{tab:complete_test2} \\
\toprule
$N$ & Vanilla & H-PINN & HO-PINN & XGB & SVR & GP \\
\midrule
\endfirsthead
\multicolumn{7}{c}{Table \thetable\ continued} \\
\toprule
$N$ & Vanilla & H-PINN & HO-PINN & XGB & SVR & GP \\
\midrule
\endhead
\midrule
\multicolumn{7}{r}{Continued on next page} \\
\endfoot
\bottomrule
\endlastfoot
\multicolumn{7}{@{}l}{\textit{Water saturation, $R^2\uparrow$ --- Interpolation}} \\*
20 & {\bfseries\boldmath $0.954\!\pm\!0.007$} & $0.954\!\pm\!0.008$ & $0.937\!\pm\!0.052$ & $0.362\!\pm\!0.241$ & $0.396\!\pm\!0.276$ & $0.541\!\pm\!0.157$ \\
50 & {\bfseries\boldmath $0.977\!\pm\!0.003$} & $0.975\!\pm\!0.004$ & $0.977\!\pm\!0.003$ & $0.632\!\pm\!0.046$ & $0.685\!\pm\!0.092$ & $0.745\!\pm\!0.089$ \\
100 & $0.987\!\pm\!0.002$ & {\bfseries\boldmath $0.988\!\pm\!0.003$} & $0.988\!\pm\!0.003$ & $0.747\!\pm\!0.052$ & $0.815\!\pm\!0.039$ & $0.876\!\pm\!0.031$ \\
200 & {\bfseries\boldmath $0.995\!\pm\!0.001$} & $0.994\!\pm\!0.001$ & $0.994\!\pm\!0.001$ & $0.836\!\pm\!0.033$ & $0.840\!\pm\!0.047$ & $0.916\!\pm\!0.017$ \\
500 & $0.998\!\pm\!0.001$ & {\bfseries\boldmath $0.998\!\pm\!0.001$} & $0.998\!\pm\!0.001$ & $0.908\!\pm\!0.025$ & $0.878\!\pm\!0.048$ & $0.960\!\pm\!0.003$ \\
1000 & $1.000\!\pm\!0.000$ & $1.000\!\pm\!0.000$ & {\bfseries\boldmath $1.000\!\pm\!0.000$} & $0.945\!\pm\!0.019$ & $0.898\!\pm\!0.035$ & $0.975\!\pm\!0.002$ \\
\addlinespace[2pt]
\multicolumn{7}{@{}l}{\textit{Water saturation, $R^2\uparrow$ --- Equation-assisted forecasting}} \\*
20 & {\bfseries\boldmath $0.917\!\pm\!0.015$} & $0.917\!\pm\!0.016$ & $0.893\!\pm\!0.079$ & $0.230\!\pm\!0.287$ & $-0.864\!\pm\!1.240$ & $0.013\!\pm\!0.392$ \\
50 & {\bfseries\boldmath $0.948\!\pm\!0.009$} & $0.947\!\pm\!0.010$ & $0.948\!\pm\!0.010$ & $0.397\!\pm\!0.136$ & $0.279\!\pm\!0.250$ & $0.486\!\pm\!0.197$ \\
100 & $0.965\!\pm\!0.005$ & $0.966\!\pm\!0.004$ & {\bfseries\boldmath $0.966\!\pm\!0.005$} & $0.503\!\pm\!0.095$ & $0.429\!\pm\!0.198$ & $0.626\!\pm\!0.150$ \\
200 & {\bfseries\boldmath $0.975\!\pm\!0.003$} & $0.974\!\pm\!0.004$ & $0.974\!\pm\!0.004$ & $0.612\!\pm\!0.080$ & $0.559\!\pm\!0.203$ & $0.739\!\pm\!0.093$ \\
500 & $0.983\!\pm\!0.004$ & {\bfseries\boldmath $0.985\!\pm\!0.003$} & $0.985\!\pm\!0.002$ & $0.686\!\pm\!0.056$ & $0.676\!\pm\!0.142$ & $0.842\!\pm\!0.014$ \\
1000 & {\bfseries\boldmath $0.993\!\pm\!0.001$} & $0.992\!\pm\!0.001$ & $0.993\!\pm\!0.001$ & $0.726\!\pm\!0.043$ & $0.730\!\pm\!0.091$ & $0.852\!\pm\!0.016$ \\
\addlinespace[2pt]
\multicolumn{7}{@{}l}{\textit{Water saturation, $\mathrm{MSE}\downarrow$ --- Interpolation}} \\*
20 & {\bfseries\boldmath $(1.1\!\pm\!0.2)10^{-3}$} & $(1.1\!\pm\!0.2)10^{-3}$ & $(1.5\!\pm\!1.3)10^{-3}$ & $(1.6\!\pm\!0.6)10^{-2}$ & $(1.5\!\pm\!0.7)10^{-2}$ & $(1.1\!\pm\!0.4)10^{-2}$ \\
50 & {\bfseries\boldmath $(5.6\!\pm\!0.7)10^{-4}$} & $(6.0\!\pm\!0.9)10^{-4}$ & $(5.7\!\pm\!0.8)10^{-4}$ & $(8.9\!\pm\!1.1)10^{-3}$ & $(7.7\!\pm\!2.2)10^{-3}$ & $(6.2\!\pm\!2.2)10^{-3}$ \\
100 & $(3.1\!\pm\!0.5)10^{-4}$ & {\bfseries\boldmath $(3.0\!\pm\!0.6)10^{-4}$} & $(3.0\!\pm\!0.7)10^{-4}$ & $(6.2\!\pm\!1.3)10^{-3}$ & $(4.5\!\pm\!1.0)10^{-3}$ & $(3.0\!\pm\!0.8)10^{-3}$ \\
200 & {\bfseries\boldmath $(1.3\!\pm\!0.3)10^{-4}$} & $(1.4\!\pm\!0.2)10^{-4}$ & $(1.4\!\pm\!0.2)10^{-4}$ & $(4.0\!\pm\!0.8)10^{-3}$ & $(3.9\!\pm\!1.1)10^{-3}$ & $(2.0\!\pm\!0.4)10^{-3}$ \\
500 & $(4.6\!\pm\!2.1)10^{-5}$ & {\bfseries\boldmath $(3.8\!\pm\!1.8)10^{-5}$} & $(4.3\!\pm\!2.5)10^{-5}$ & $(2.2\!\pm\!0.6)10^{-3}$ & $(3.0\!\pm\!1.2)10^{-3}$ & $(9.8\!\pm\!0.8)10^{-4}$ \\
1000 & $(8.2\!\pm\!2.5)10^{-6}$ & $(8.3\!\pm\!3.1)10^{-6}$ & {\bfseries\boldmath $(8.1\!\pm\!2.5)10^{-6}$} & $(1.3\!\pm\!0.5)10^{-3}$ & $(2.5\!\pm\!0.9)10^{-3}$ & $(6.1\!\pm\!0.5)10^{-4}$ \\
\addlinespace[2pt]
\multicolumn{7}{@{}l}{\textit{Water saturation, $\mathrm{MSE}\downarrow$ --- Equation-assisted forecasting}} \\*
20 & {\bfseries\boldmath $(2.9\!\pm\!0.5)10^{-3}$} & $(2.9\!\pm\!0.6)10^{-3}$ & $(3.7\!\pm\!2.7)10^{-3}$ & $(2.7\!\pm\!1.0)10^{-2}$ & $(6.4\!\pm\!4.3)10^{-2}$ & $(3.4\!\pm\!1.4)10^{-2}$ \\
50 & {\bfseries\boldmath $(1.8\!\pm\!0.3)10^{-3}$} & $(1.8\!\pm\!0.4)10^{-3}$ & $(1.8\!\pm\!0.3)10^{-3}$ & $(2.1\!\pm\!0.5)10^{-2}$ & $(2.5\!\pm\!0.9)10^{-2}$ & $(1.8\!\pm\!0.7)10^{-2}$ \\
100 & $(1.2\!\pm\!0.2)10^{-3}$ & $(1.2\!\pm\!0.1)10^{-3}$ & {\bfseries\boldmath $(1.2\!\pm\!0.2)10^{-3}$} & $(1.7\!\pm\!0.3)10^{-2}$ & $(2.0\!\pm\!0.7)10^{-2}$ & $(1.3\!\pm\!0.5)10^{-2}$ \\
200 & {\bfseries\boldmath $(8.8\!\pm\!1.2)10^{-4}$} & $(9.0\!\pm\!1.4)10^{-4}$ & $(9.0\!\pm\!1.4)10^{-4}$ & $(1.3\!\pm\!0.3)10^{-2}$ & $(1.5\!\pm\!0.7)10^{-2}$ & $(9.0\!\pm\!3.2)10^{-3}$ \\
500 & $(5.8\!\pm\!1.3)10^{-4}$ & {\bfseries\boldmath $(5.0\!\pm\!0.9)10^{-4}$} & $(5.1\!\pm\!0.6)10^{-4}$ & $(1.1\!\pm\!0.2)10^{-2}$ & $(1.1\!\pm\!0.5)10^{-2}$ & $(5.5\!\pm\!0.5)10^{-3}$ \\
1000 & {\bfseries\boldmath $(2.5\!\pm\!0.3)10^{-4}$} & $(2.8\!\pm\!0.3)10^{-4}$ & $(2.6\!\pm\!0.3)10^{-4}$ & $(9.5\!\pm\!1.5)10^{-3}$ & $(9.3\!\pm\!3.1)10^{-3}$ & $(5.1\!\pm\!0.5)10^{-3}$ \\
\end{longtable}
\endgroup

\begingroup
\fontsize{8}{10}\selectfont
\setlength{\tabcolsep}{0.5pt}
\renewcommand{\arraystretch}{0.98}
\setlength{\LTpre}{5pt}
\setlength{\LTpost}{5pt}
\begin{longtable}{@{}r@{\hspace{5pt}}cccccc@{}}
\caption{Test 3 water saturation, all budgets (mean $\pm$ sample std over ten seeds). Bold marks the best unrounded mean. MSE entries use $(a\pm b)10^e$.}
\label{tab:complete_test3} \\
\toprule
$N$ & Vanilla & H-PINN & HO-PINN & XGB & SVR & GP \\
\midrule
\endfirsthead
\multicolumn{7}{c}{Table \thetable\ continued} \\
\toprule
$N$ & Vanilla & H-PINN & HO-PINN & XGB & SVR & GP \\
\midrule
\endhead
\midrule
\multicolumn{7}{r}{Continued on next page} \\
\endfoot
\bottomrule
\endlastfoot
\multicolumn{7}{@{}l}{\textit{Water saturation, $R^2\uparrow$ --- Interpolation}} \\*
20 & {\bfseries\boldmath $0.755\!\pm\!0.018$} & $0.461\!\pm\!0.360$ & $0.567\!\pm\!0.312$ & $0.019\!\pm\!0.350$ & $0.084\!\pm\!0.482$ & $0.265\!\pm\!0.401$ \\
50 & {\bfseries\boldmath $0.847\!\pm\!0.015$} & $0.841\!\pm\!0.015$ & $0.842\!\pm\!0.010$ & $0.344\!\pm\!0.302$ & $0.377\!\pm\!0.304$ & $0.544\!\pm\!0.247$ \\
100 & {\bfseries\boldmath $0.908\!\pm\!0.023$} & $0.905\!\pm\!0.020$ & $0.905\!\pm\!0.020$ & $0.604\!\pm\!0.122$ & $0.596\!\pm\!0.171$ & $0.748\!\pm\!0.211$ \\
200 & {\bfseries\boldmath $0.947\!\pm\!0.009$} & $0.946\!\pm\!0.010$ & $0.946\!\pm\!0.010$ & $0.770\!\pm\!0.067$ & $0.711\!\pm\!0.049$ & $0.891\!\pm\!0.019$ \\
500 & $0.979\!\pm\!0.005$ & $0.980\!\pm\!0.005$ & {\bfseries\boldmath $0.980\!\pm\!0.005$} & $0.888\!\pm\!0.018$ & $0.779\!\pm\!0.063$ & $0.949\!\pm\!0.006$ \\
1000 & $0.990\!\pm\!0.002$ & $0.991\!\pm\!0.002$ & {\bfseries\boldmath $0.991\!\pm\!0.002$} & $0.931\!\pm\!0.019$ & $0.766\!\pm\!0.065$ & $0.969\!\pm\!0.004$ \\
\addlinespace[2pt]
\multicolumn{7}{@{}l}{\textit{Water saturation, $R^2\uparrow$ --- Equation-assisted forecasting}} \\*
20 & {\bfseries\boldmath $0.642\!\pm\!0.042$} & $0.334\!\pm\!0.401$ & $0.445\!\pm\!0.365$ & $-0.089\!\pm\!0.269$ & $-2.148\!\pm\!4.703$ & $0.084\!\pm\!0.284$ \\
50 & {\bfseries\boldmath $0.768\!\pm\!0.033$} & $0.762\!\pm\!0.026$ & $0.759\!\pm\!0.025$ & $0.172\!\pm\!0.394$ & $-1.011\!\pm\!2.289$ & $0.293\!\pm\!0.269$ \\
100 & {\bfseries\boldmath $0.857\!\pm\!0.029$} & $0.855\!\pm\!0.026$ & $0.853\!\pm\!0.026$ & $0.440\!\pm\!0.137$ & $-0.636\!\pm\!2.185$ & $0.551\!\pm\!0.214$ \\
200 & $0.915\!\pm\!0.015$ & $0.915\!\pm\!0.015$ & {\bfseries\boldmath $0.916\!\pm\!0.015$} & $0.614\!\pm\!0.050$ & $-0.389\!\pm\!1.808$ & $0.678\!\pm\!0.131$ \\
500 & $0.954\!\pm\!0.005$ & $0.956\!\pm\!0.005$ & {\bfseries\boldmath $0.957\!\pm\!0.006$} & $0.740\!\pm\!0.043$ & $0.231\!\pm\!0.487$ & $0.802\!\pm\!0.036$ \\
1000 & $0.972\!\pm\!0.004$ & $0.974\!\pm\!0.003$ & {\bfseries\boldmath $0.974\!\pm\!0.003$} & $0.809\!\pm\!0.032$ & $0.422\!\pm\!0.186$ & $0.838\!\pm\!0.031$ \\
\addlinespace[2pt]
\multicolumn{7}{@{}l}{\textit{Water saturation, $\mathrm{MSE}\downarrow$ --- Interpolation}} \\*
20 & {\bfseries\boldmath $(3.7\!\pm\!0.3)10^{-3}$} & $(8.1\!\pm\!5.4)10^{-3}$ & $(6.5\!\pm\!4.7)10^{-3}$ & $(1.5\!\pm\!0.5)10^{-2}$ & $(1.4\!\pm\!0.7)10^{-2}$ & $(1.1\!\pm\!0.6)10^{-2}$ \\
50 & {\bfseries\boldmath $(2.3\!\pm\!0.2)10^{-3}$} & $(2.4\!\pm\!0.2)10^{-3}$ & $(2.4\!\pm\!0.2)10^{-3}$ & $(9.8\!\pm\!4.5)10^{-3}$ & $(9.3\!\pm\!4.6)10^{-3}$ & $(6.8\!\pm\!3.7)10^{-3}$ \\
100 & {\bfseries\boldmath $(1.4\!\pm\!0.3)10^{-3}$} & $(1.4\!\pm\!0.3)10^{-3}$ & $(1.4\!\pm\!0.3)10^{-3}$ & $(5.9\!\pm\!1.8)10^{-3}$ & $(6.1\!\pm\!2.6)10^{-3}$ & $(3.8\!\pm\!3.2)10^{-3}$ \\
200 & {\bfseries\boldmath $(8.0\!\pm\!1.4)10^{-4}$} & $(8.0\!\pm\!1.6)10^{-4}$ & $(8.0\!\pm\!1.5)10^{-4}$ & $(3.4\!\pm\!1.0)10^{-3}$ & $(4.3\!\pm\!0.7)10^{-3}$ & $(1.6\!\pm\!0.3)10^{-3}$ \\
500 & $(3.2\!\pm\!0.7)10^{-4}$ & $(3.0\!\pm\!0.7)10^{-4}$ & {\bfseries\boldmath $(3.0\!\pm\!0.7)10^{-4}$} & $(1.7\!\pm\!0.3)10^{-3}$ & $(3.3\!\pm\!0.9)10^{-3}$ & $(7.6\!\pm\!0.9)10^{-4}$ \\
1000 & $(1.5\!\pm\!0.4)10^{-4}$ & $(1.4\!\pm\!0.3)10^{-4}$ & {\bfseries\boldmath $(1.4\!\pm\!0.3)10^{-4}$} & $(1.0\!\pm\!0.3)10^{-3}$ & $(3.5\!\pm\!1.0)10^{-3}$ & $(4.6\!\pm\!0.6)10^{-4}$ \\
\addlinespace[2pt]
\multicolumn{7}{@{}l}{\textit{Water saturation, $\mathrm{MSE}\downarrow$ --- Equation-assisted forecasting}} \\*
20 & {\bfseries\boldmath $(1.0\!\pm\!0.1)10^{-2}$} & $(1.9\!\pm\!1.2)10^{-2}$ & $(1.6\!\pm\!1.1)10^{-2}$ & $(3.1\!\pm\!0.8)10^{-2}$ & $(9.1\!\pm\!13.5)10^{-2}$ & $(2.6\!\pm\!0.8)10^{-2}$ \\
50 & {\bfseries\boldmath $(6.7\!\pm\!0.9)10^{-3}$} & $(6.8\!\pm\!0.8)10^{-3}$ & $(6.9\!\pm\!0.7)10^{-3}$ & $(2.4\!\pm\!1.1)10^{-2}$ & $(5.8\!\pm\!6.6)10^{-2}$ & $(2.0\!\pm\!0.8)10^{-2}$ \\
100 & {\bfseries\boldmath $(4.1\!\pm\!0.8)10^{-3}$} & $(4.2\!\pm\!0.7)10^{-3}$ & $(4.2\!\pm\!0.7)10^{-3}$ & $(1.6\!\pm\!0.4)10^{-2}$ & $(4.7\!\pm\!6.3)10^{-2}$ & $(1.3\!\pm\!0.6)10^{-2}$ \\
200 & $(2.5\!\pm\!0.4)10^{-3}$ & $(2.4\!\pm\!0.4)10^{-3}$ & {\bfseries\boldmath $(2.4\!\pm\!0.4)10^{-3}$} & $(1.1\!\pm\!0.1)10^{-2}$ & $(4.0\!\pm\!5.2)10^{-2}$ & $(9.3\!\pm\!3.8)10^{-3}$ \\
500 & $(1.3\!\pm\!0.2)10^{-3}$ & $(1.3\!\pm\!0.2)10^{-3}$ & {\bfseries\boldmath $(1.2\!\pm\!0.2)10^{-3}$} & $(7.5\!\pm\!1.2)10^{-3}$ & $(2.2\!\pm\!1.4)10^{-2}$ & $(5.7\!\pm\!1.0)10^{-3}$ \\
1000 & $(8.0\!\pm\!1.0)10^{-4}$ & $(7.5\!\pm\!0.8)10^{-4}$ & {\bfseries\boldmath $(7.4\!\pm\!0.9)10^{-4}$} & $(5.5\!\pm\!0.9)10^{-3}$ & $(1.7\!\pm\!0.5)10^{-2}$ & $(4.7\!\pm\!0.9)10^{-3}$ \\
\end{longtable}
\endgroup

\begingroup
\fontsize{8}{10}\selectfont
\setlength{\tabcolsep}{0.5pt}
\renewcommand{\arraystretch}{0.98}
\setlength{\LTpre}{5pt}
\setlength{\LTpost}{5pt}
\begin{longtable}{@{}r@{\hspace{5pt}}cccccc@{}}
\caption{Test 4 water saturation, all budgets (mean $\pm$ sample std over ten seeds). Bold marks the best unrounded mean. MSE entries use $(a\pm b)10^e$.}
\label{tab:complete_test4} \\
\toprule
$N$ & Vanilla & H-PINN & HO-PINN & XGB & SVR & GP \\
\midrule
\endfirsthead
\multicolumn{7}{c}{Table \thetable\ continued} \\
\toprule
$N$ & Vanilla & H-PINN & HO-PINN & XGB & SVR & GP \\
\midrule
\endhead
\midrule
\multicolumn{7}{r}{Continued on next page} \\
\endfoot
\bottomrule
\endlastfoot
\multicolumn{7}{@{}l}{\textit{Water saturation, $R^2\uparrow$ --- Interpolation}} \\*
20 & $0.495\!\pm\!0.251$ & {\bfseries\boldmath $0.522\!\pm\!0.305$} & $0.426\!\pm\!0.358$ & $-0.058\!\pm\!0.268$ & $-0.187\!\pm\!0.294$ & $-0.077\!\pm\!0.339$ \\
50 & $0.777\!\pm\!0.100$ & {\bfseries\boldmath $0.807\!\pm\!0.015$} & $0.779\!\pm\!0.109$ & $0.182\!\pm\!0.160$ & $0.083\!\pm\!0.109$ & $0.282\!\pm\!0.192$ \\
100 & $0.864\!\pm\!0.013$ & {\bfseries\boldmath $0.867\!\pm\!0.013$} & $0.865\!\pm\!0.013$ & $0.419\!\pm\!0.065$ & $0.230\!\pm\!0.094$ & $0.575\!\pm\!0.210$ \\
200 & $0.912\!\pm\!0.010$ & $0.913\!\pm\!0.013$ & {\bfseries\boldmath $0.915\!\pm\!0.009$} & $0.591\!\pm\!0.059$ & $0.263\!\pm\!0.124$ & $0.783\!\pm\!0.090$ \\
500 & $0.963\!\pm\!0.006$ & $0.962\!\pm\!0.006$ & {\bfseries\boldmath $0.963\!\pm\!0.005$} & $0.738\!\pm\!0.042$ & $0.450\!\pm\!0.074$ & $0.885\!\pm\!0.021$ \\
1000 & {\bfseries\boldmath $0.983\!\pm\!0.005$} & $0.983\!\pm\!0.004$ & $0.982\!\pm\!0.004$ & $0.831\!\pm\!0.018$ & $0.527\!\pm\!0.032$ & $0.933\!\pm\!0.006$ \\
\addlinespace[2pt]
\multicolumn{7}{@{}l}{\textit{Water saturation, $R^2\uparrow$ --- Equation-assisted forecasting}} \\*
20 & {\bfseries\boldmath $0.558\!\pm\!0.207$} & $0.550\!\pm\!0.360$ & $0.435\!\pm\!0.454$ & $-0.124\!\pm\!0.314$ & $-0.325\!\pm\!0.364$ & $-0.084\!\pm\!0.280$ \\
50 & $0.794\!\pm\!0.059$ & {\bfseries\boldmath $0.815\!\pm\!0.011$} & $0.799\!\pm\!0.067$ & $0.198\!\pm\!0.269$ & $0.019\!\pm\!0.180$ & $0.216\!\pm\!0.214$ \\
100 & $0.864\!\pm\!0.012$ & {\bfseries\boldmath $0.866\!\pm\!0.013$} & $0.863\!\pm\!0.015$ & $0.405\!\pm\!0.064$ & $0.231\!\pm\!0.120$ & $0.529\!\pm\!0.189$ \\
200 & $0.905\!\pm\!0.011$ & $0.903\!\pm\!0.015$ & {\bfseries\boldmath $0.907\!\pm\!0.011$} & $0.521\!\pm\!0.100$ & $0.258\!\pm\!0.195$ & $0.704\!\pm\!0.099$ \\
500 & $0.951\!\pm\!0.007$ & $0.950\!\pm\!0.006$ & {\bfseries\boldmath $0.951\!\pm\!0.006$} & $0.663\!\pm\!0.040$ & $0.405\!\pm\!0.118$ & $0.797\!\pm\!0.048$ \\
1000 & {\bfseries\boldmath $0.972\!\pm\!0.004$} & $0.971\!\pm\!0.004$ & $0.971\!\pm\!0.004$ & $0.751\!\pm\!0.025$ & $0.476\!\pm\!0.055$ & $0.856\!\pm\!0.019$ \\
\addlinespace[2pt]
\multicolumn{7}{@{}l}{\textit{Water saturation, $\mathrm{MSE}\downarrow$ --- Interpolation}} \\*
20 & $(6.0\!\pm\!3.0)10^{-3}$ & {\bfseries\boldmath $(5.7\!\pm\!3.6)10^{-3}$} & $(6.8\!\pm\!4.3)10^{-3}$ & $(1.3\!\pm\!0.3)10^{-2}$ & $(1.4\!\pm\!0.3)10^{-2}$ & $(1.3\!\pm\!0.4)10^{-2}$ \\
50 & $(2.6\!\pm\!1.2)10^{-3}$ & {\bfseries\boldmath $(2.3\!\pm\!0.2)10^{-3}$} & $(2.6\!\pm\!1.3)10^{-3}$ & $(9.7\!\pm\!1.9)10^{-3}$ & $(1.1\!\pm\!0.1)10^{-2}$ & $(8.5\!\pm\!2.3)10^{-3}$ \\
100 & $(1.6\!\pm\!0.2)10^{-3}$ & {\bfseries\boldmath $(1.6\!\pm\!0.2)10^{-3}$} & $(1.6\!\pm\!0.2)10^{-3}$ & $(6.9\!\pm\!0.8)10^{-3}$ & $(9.2\!\pm\!1.1)10^{-3}$ & $(5.1\!\pm\!2.5)10^{-3}$ \\
200 & $(1.0\!\pm\!0.1)10^{-3}$ & $(1.0\!\pm\!0.2)10^{-3}$ & {\bfseries\boldmath $(1.0\!\pm\!0.1)10^{-3}$} & $(4.9\!\pm\!0.7)10^{-3}$ & $(8.8\!\pm\!1.5)10^{-3}$ & $(2.6\!\pm\!1.1)10^{-3}$ \\
500 & $(4.5\!\pm\!0.7)10^{-4}$ & $(4.6\!\pm\!0.7)10^{-4}$ & {\bfseries\boldmath $(4.4\!\pm\!0.5)10^{-4}$} & $(3.1\!\pm\!0.5)10^{-3}$ & $(6.5\!\pm\!0.9)10^{-3}$ & $(1.4\!\pm\!0.3)10^{-3}$ \\
1000 & {\bfseries\boldmath $(2.1\!\pm\!0.5)10^{-4}$} & $(2.1\!\pm\!0.5)10^{-4}$ & $(2.1\!\pm\!0.5)10^{-4}$ & $(2.0\!\pm\!0.2)10^{-3}$ & $(5.6\!\pm\!0.4)10^{-3}$ & $(8.0\!\pm\!0.8)10^{-4}$ \\
\addlinespace[2pt]
\multicolumn{7}{@{}l}{\textit{Water saturation, $\mathrm{MSE}\downarrow$ --- Equation-assisted forecasting}} \\*
20 & {\bfseries\boldmath $(1.2\!\pm\!0.5)10^{-2}$} & $(1.2\!\pm\!1.0)10^{-2}$ & $(1.5\!\pm\!1.2)10^{-2}$ & $(3.0\!\pm\!0.8)10^{-2}$ & $(3.5\!\pm\!1.0)10^{-2}$ & $(2.9\!\pm\!0.7)10^{-2}$ \\
50 & $(5.4\!\pm\!1.6)10^{-3}$ & {\bfseries\boldmath $(4.9\!\pm\!0.3)10^{-3}$} & $(5.3\!\pm\!1.8)10^{-3}$ & $(2.1\!\pm\!0.7)10^{-2}$ & $(2.6\!\pm\!0.5)10^{-2}$ & $(2.1\!\pm\!0.6)10^{-2}$ \\
100 & $(3.6\!\pm\!0.3)10^{-3}$ & {\bfseries\boldmath $(3.5\!\pm\!0.4)10^{-3}$} & $(3.6\!\pm\!0.4)10^{-3}$ & $(1.6\!\pm\!0.2)10^{-2}$ & $(2.0\!\pm\!0.3)10^{-2}$ & $(1.2\!\pm\!0.5)10^{-2}$ \\
200 & $(2.5\!\pm\!0.3)10^{-3}$ & $(2.6\!\pm\!0.4)10^{-3}$ & {\bfseries\boldmath $(2.5\!\pm\!0.3)10^{-3}$} & $(1.3\!\pm\!0.3)10^{-2}$ & $(2.0\!\pm\!0.5)10^{-2}$ & $(7.8\!\pm\!2.6)10^{-3}$ \\
500 & $(1.3\!\pm\!0.2)10^{-3}$ & $(1.3\!\pm\!0.2)10^{-3}$ & {\bfseries\boldmath $(1.3\!\pm\!0.2)10^{-3}$} & $(8.9\!\pm\!1.1)10^{-3}$ & $(1.6\!\pm\!0.3)10^{-2}$ & $(5.4\!\pm\!1.3)10^{-3}$ \\
1000 & {\bfseries\boldmath $(7.4\!\pm\!1.2)10^{-4}$} & $(7.6\!\pm\!1.1)10^{-4}$ & $(7.8\!\pm\!1.0)10^{-4}$ & $(6.6\!\pm\!0.7)10^{-3}$ & $(1.4\!\pm\!0.1)10^{-2}$ & $(3.8\!\pm\!0.5)10^{-3}$ \\
\end{longtable}
\endgroup

\endgroup
\FloatBarrier

\end{document}